\documentclass[draftclsnofoot]{IEEEtran}
\usepackage{graphicx}
\usepackage{epstopdf}

\usepackage{subfigure}
\usepackage{booktabs}
\usepackage{caption}
\usepackage{color}
\usepackage{bm}
\usepackage{CJK}
\usepackage{indentfirst}
\usepackage{amsmath}
\usepackage{amssymb}
\usepackage{float}
\usepackage{multirow}
\usepackage{algorithm}
\usepackage{algpseudocode}
\usepackage{amsmath}
\usepackage{flushend}
\usepackage{setspace}
\usepackage{geometry}
\newtheorem{lemma}{Lemma}

\begin{document}
\title{{Channel Customization for Joint Tx-RISs-Rx Design in Hybrid mmWave \textcolor{black}{Systems}}}
\author{\IEEEauthorblockN{Weicong~Chen, Chao-Kai Wen, Xiao Li, and~Shi~Jin}\\
\thanks{This paper was presented in part at IEEE ICC, May 2022 \cite{ICC-BPA}.}
\thanks{{Weicong Chen, Xiao Li, and Shi Jin are with the National Mobile Communications
Research Laboratory, Southeast University, Nanjing 210096, China (e-mail: cwc@seu.edu.cn; li\_xiao@seu.edu.cn; jinshi@seu.edu.cn).} }
\thanks{Chao-Kai Wen is with the Institute of Communications Engineering, National Sun Yat-sen University, Kaohsiung 80424, Taiwan. (e-mail: chaokai.wen@mail.nsysu.edu.tw).}
}
\maketitle
\begin{abstract}
In strong line-of-sight millimeter-wave (mmWave) wireless systems, the rank-deficient channel severely hampers spatial multiplexing. To address this inherent deficiency, multiple reconfigurable-intelligent-surfaces (RISs) are introduced in this study to customize the wireless channel. Utilizing the RIS to reshape electromagnetic waves, we theoretically show that a favorable channel with an arbitrary tunable rank and a minimized truncated condition number can be established by elaborately designing the placement and reflection matrix of RISs. Different from existing works on multi-RISs, the number of elements needed for each RIS to combat the path loss and the limited phase control is also considered. On the basis of the proposed channel customization, a joint transmitter-RISs-receiver (Tx-RISs-Rx) design under a hybrid mmWave system is investigated to maximize the spectral efficiency. Using the proposed scheme, the optimal singular value decomposition-based hybrid beamforming at the Tx and Rx can be obtained without matrix decomposition for the digital and analog beamforming. The bottoms of the sub-channel mode in the water-filling algorithm, which are conventionally uncontrollable, are proven to be independently adjustable by RISs. Moreover, the transmit power required for realizing multi-stream transmission is derived. Numerical results are presented to verify our theoretical analysis and exhibit substantial gains over systems without RISs.
\end{abstract}

\section{Introduction}
Since Shannon defined the maximum information transfer rate of a communication channel \cite{Shannon}, the entire communication industry has been constantly improving and approaching the Shannon limit. According to Shannon's definition, the maximum information transfer rate, or channel capacity, depends on the signal-to-noise-ratio (SNR) and bandwidth of the channel. Thus, improving the SNR and increasing the bandwidth are two straightforward approaches to enlarge the capacity. However, logarithmic scaling indicates a rapidly diminishing gain obtained from SNR, and the commercial bandwidth is limited and expensive. Consequently, the multichannel triggered  by the multiple-input-multiple-output (MIMO) technology was explored to pave a new way for capacity promotion \cite{MIMO_capa}. In MIMO systems, the capacity grows linearly with the number of multichannels determined by the rank of the channel \cite{MIMO_OFDM}. Nevertheless, when the troika (SNR, bandwidth, and multichannel) steps into millimeter-wave (mmWave) MIMO systems, new dilemmas emerge.\par

{MmWave has been regarded as a promising advancement for future wireless communication systems. In contrast to the sub-6 GHz where the spectrum is extremely scarce, mmWave spectra above 20 GHz can provide multiple frequency bands with bandwidth up to 1--2 GHz, which offers extreme broadband capacity \cite{D. Soldani}\cite{multiband}. Although promising, the coverage of transmission in mmWave spectra shrinks due to severe path loss \cite{R. Baldemair}. Equipping a large-scale antenna array \cite{M. R. Akdeniz}--\cite{Z. Xiao} to provide more array gains is a key solution to improve the transmission rate and communication distance.} Considering that the upper limit of channel rank is imposed by the minimum number of antennas, the high SNR, broad bandwidth, and massive antennas in mmWave MIMO systems are supposed to enhance the capacity significantly. However, channel rank is also constrained by the number of propagation paths. The combination of high gain beams and sparse channel in mmWave systems severely reduces the multichannel number, \textcolor{black}{which may plummet to $1$ in the strong far-field  line of sight (LoS) scenario} regardless of the increasing antennas. The crux of spatial multiplexing provided by the multichannel lies in the channel that has been conventionally modeled as an uncontrollable exogenous entity. This iron law for the uncontrollable channel is broken by the burgeoning reconfigurable intelligent surface (RIS).\par

RIS, also known as passive large intelligent surface (LIS) \cite{Y. Han-LIS} or intelligent reflecting surface (IRS) \cite{Q. Wu-IRS}, has been envisaged to realize the so-called smart radio environment \cite{smart-radio} that has become part of the system design parameters. It has been used to transform the channel for achieving multifarious tasks in various scenarios \cite{App-mmWave-1}--\cite{opt_ref_5}. To enhance the coverage in mmWave systems, \cite{App-mmWave-1} and \cite{App-mmWave-2} introduced RIS to alleviate the significant path loss and severe blockage, respectively. In \cite{W. Tang_1} and \cite{W. Tang_2}, RISs were used for beamforming and broadcasting, and the corresponding large-scale fading models were provided together with measurement verifications. In \cite{App_scerecy}, RIS was utilized to create friendly multipaths for enhancing the secrecy rate of directional modulation. In multi-user scenarios, the maximization problem for the sum rate \cite{opt_ref_1}, the weighted sum rate \cite{opt_ref_2}, the energy efficiency \cite{App_EE}, the minimum signal-to-interference-plus-noise \cite{opt_ref_3}, the minimization problem for the average transmit power \cite{opt_ref_4}, and the maximum learning error \cite{opt_ref_5} were investigated by optimizing the reflection matrix of the RIS.\par

Most of the aforementioned works mainly focused on solving optimization problems and merely illustrated how the RIS can reshape the channel. A joint transceiver and RIS design that can realize a favorable propagation environment with a small channel matrix condition number was reported in \cite{joint_Dai}. \textcolor{black}{By deploying multiple RISs in the channel, \cite{X. Yang}--\cite{Rev-R2} demonstrated that channel rank can be improved.} Meanwhile, \cite{max_erank} examined the maximization of the effective rank and the minimum singular value of the RIS-augmented channel by using gradient-based optimization. With RIS placed within a random environment, the authors of \cite{NE} experimentally demonstrated that disorder environments can be tuned to achieve  optimal channel diversity  by physically shaping the propagation medium itself. \textcolor{black}{In a near-field scenario wherein only the LoS link cascaded by RIS exists, \cite{Rev-R1} investigated \textcolor{black}{the transmitter-RIS (Tx-RIS) and RIS-receiver (RIS-Rx)} distance pair to fully utilize multiplexing communication. In \cite{Rev-R3}, wherein the double-RIS-assisted MIMO communication under LoS channels was considered, the authors determined that when the Tx and the Rx have different array response vectors for varying RISs, the MIMO channel is of rank-$2$; otherwise, it is of rank-$1$. } Unfortunately, no existing works have theoretically explained how the reflection phases of the RIS determine the characteristics of the channel, such as its condition number, rank, or effective rank.\par

Motivated by the intrinsic rank deficiency in \textcolor{black}{the strong far-field  LoS mmWave channel} and the potential of RISs in reshaping the channel, we study the channel customization to configure a favorable wireless propagation environment for a joint transmitter-RISs-receiver design in the multiple RISs-assisted mmWave system. The contributions of this work are summarized as follows.
\begin{itemize}
  \item A channel customization method is proposed to modify singular values of the composite channel between the Tx and Rx in three strong LoS mmWave scenarios with different antenna configurations. We demonstrate that by carefully designing the reflection matrix, set segmentation, and deployment location of the RISs, channel rank can be arbitrarily customized according to the number of data streams, and a well-conditioned channel with the minimum truncated condition number can be established.
  \item A joint Tx-RISs-Rx design scheme is developed under the hybrid mmWave architecture, which aims to maximize the downlink spectral efficiency (SE) by jointly devising the hybrid beamforming at the Tx and Rx, and the reflection phases of RISs. On the basis of the proposed channel customization, the optimal singular value decomposition (SVD) based hybrid beamforming can be easily separated into digital and analog beamforming without any matrix decomposition algorithms.
  \item The required transmit power for realizing $s$-stream transmission is derived. Different from the conventional water-filling power allocation, our channel customization method provides a novel way to change the bottom of the sub-channel mode individually by adjusting the tunable singular values. In such conditions, the transmit power threshold that enables the water level to cover $s$ sub-channels modes is determined by $s$ and the concrete design of the reflection matrix of RISs.
  \item Numerical results are presented to illustrate the effectiveness of the proposed channel customization and joint Tx-RISs-Rx design. With the help of RISs, a favorable channel with an arbitrary adjustable channel rank and a small truncated condition number can be customized. Furthermore, the joint Tx-RISs-Rx transmission scheme reaps substantial gains compared with conventional systems without RISs.
\end{itemize}

The rest of this paper is organized as follows. Section \ref{sec-2} introduces the system model. Then, the proposed channel customization and joint Tx-RISs-Rx design are presented in Sections \ref{sec-3} and \ref{sec-4}, respectively. Afterward, Section \ref{sec-5} provides the numerical results for our proposal. Finally, Section \ref{sec-6} concludes this study.

\emph{Notations}: The vector and matrix are denoted by the lowercase and uppercase of a letter, respectively. ${\rm tr}(\cdot)$ calculates the trace of a matrix. The transpose and conjugated-transpose operations are represented by superscripts $(\cdot)^T$ and $(\cdot)^H$, respectively. The superscripts $(\cdot)^{\rm A}$ and $(\cdot)^{\rm D}$ indicate parameters that are associated with the angle-of-arrival (AoA) and angle-of-departure (AoD), respectively. \textcolor{black}{The subscripts $(\cdot)_{\rm D}$, $(\cdot)_{\rm T}$, $(\cdot)_{\rm R}$, and $(\cdot)_{\rm S}$ stand for Direct, Tx, Rx, and Surface (RIS), respectively}. $|\cdot|$ and $\|\cdot\|_F$ are used to indicate the absolute value and Euclidean norm, respectively. $\lceil\cdot\rceil$ is the integer ceiling. ${\rm blkdiag}\{{\bf X}_1,{\bf X}_2,\cdots,{\bf X}_N\}$ represents a block diagonal matrix with diagonal matrices ${\bf X}_i$, $i=1,\cdots,N$, while ${\rm diag}(a_1,a_2,\cdots,a_N)$ indicates a diagonal matrix with elements $a_i$, $i=1,\cdots,N$. The Kronecker product is denoted by $\otimes$.

\section{System Model}\label{sec-2}

\begin{figure}[!t]
\centering
    \includegraphics[width=0.5\textwidth]{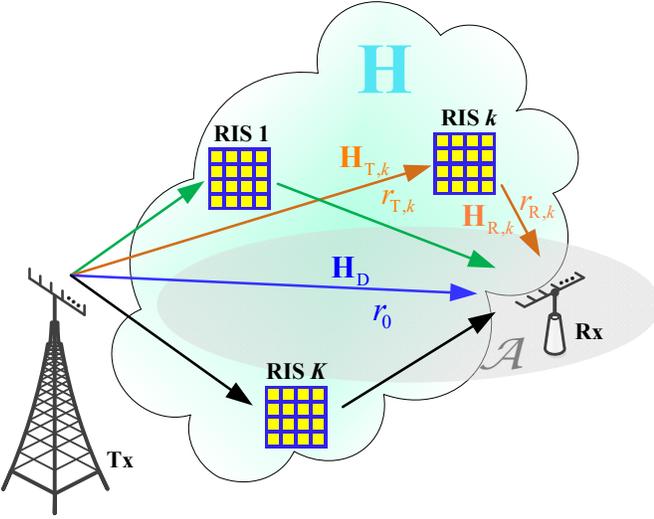}
\caption{Multiple RISs-assisted hybrid mmWave systems}
\label{Fig.system_model}
\end{figure}

We consider a multiple RISs-assisted narrowband hybrid mmWave MIMO communication system, as shown in Fig. \ref{Fig.system_model}, where $K$ RISs assist a multi-antenna Tx to serve a multi-antenna Rx. The uniform linear array\footnote{This work can be easily extended to the scenarios where the Tx and Rx are equipped with the uniform planar array (UPA). However, additional angle parameters will be introduced with the UPA. For clear and concise description of our main idea, we consider the ULA to omit redundant expressions.} (ULA) with element spacing being $d_{\rm T}$ and $d_{\rm R}$ is used at the Tx and Rx, respectively. The Tx is equipped with $N_{\rm T}$ antennas and $K_{\rm T}$ \textcolor{black}{radio frequency (RF)} chains, and the Rx is equipped with \textcolor{black}{$N_{\rm R}$ antennas and $K_{\rm R}$ RF chains}, where $K_{\rm T}<N_{\rm T}$ and $K_{\rm R}<N_{\rm R}$. Considering that the distances between Tx--RIS and RIS--Rx are different among RISs, we suggest that RISs be equipped with sufficient elements \textcolor{black}{(typically much larger than $N_{\rm T}$ and $N_{\rm R}$)} to combat diverse path losses. Therefore, we assume that an $N_{{\rm S},k}$($=N_{{\rm S,v},k}\times N_{{\rm S,h},k}$)-element UPA is equipped at RIS $k$, $k\in {\mathcal K}=\{1, 2, \cdots, K\}$. The UPA for RIS $k$ consists of $N_{{\rm S,v},k}$ rows and $N_{{\rm S,h},k}$ columns with all spacings  being $d_{\rm S}$. In practice, the continuous and element-level control of the RIS reflection phase is costly to implement. For ease of practical implementation, we consider the column-control \cite{AngleReci} with the reflection phase taken only from a finite number of discrete values. The column-controlled RIS can be regarded as ULA thus reducing the system to two-dimension. Without loss of generality, we omit the height of the Tx, RISs, and Rx by assuming that they are at the same altitude. For system deployment, the Tx is located at ${{\bf{e}}_{\rm{T}}} = {[ {{x_{\rm{T}}},{y_{\rm{T}}}} ]^T}$ in the Cartesian coordinates, the Rx's location ${{\bf{e}}_{\rm{R}}} = {[ {{x_{\rm{R}}},{y_{\rm{R}}}} ]^T}$ is randomly distributed in coverage area ${\mathcal A}$, and the position of RIS $k$, ${{\bf{e}}_{{\rm{S,}}k}} = {[ {{x_{{\rm{S,}}k}},{y_{{\rm{S,}}k}}} ]^T}$, can be flexibly designed to customize the channel. \textcolor{black}{The LoS distances for the Tx-RIS $k$, Rx--RIS $k$, and Tx-Rx channels are given by ${r_{{\rm T},k}}=\|{{\bf{e}}_{{\rm{T}}}}-{{\bf{e}}_{{\rm{S,}}k}}\|$, ${r_{{\rm R},k}}=\|{{\bf{e}}_{{\rm{R}}}}-{{\bf{e}}_{{\rm{S,}}k}}\|$, and ${r_0}=\|{{\bf{e}}_{{\rm{T}}}}-{{\bf{e}}_{{\rm{R}}}}\|$, respectively.}\par

\par

In the downlink, the received signal at the Rx is given by
\begin{equation}\label{eq-y-origanl}
  {\bf{y}} = {\bf{W}}_{{\rm{BB}}}^H{\bf{W}}_{{\rm{RF}}}^H{\bf{H}}{{\bf{F}}_{{\rm{RF}}}}{{\bf{F}}_{{\rm{BB}}}}{\bf{x}} + {\bf{W}}_{{\rm{BB}}}^H{\bf{W}}_{{\rm{RF}}}^H{\bf{n}},
\end{equation}
where ${\bf{x}} = {[ {{x_1},{x_2}, \cdots ,{x_s}} ]^T} \in {\mathbb C}{^{s \times 1}}$ satisfying ${\mathbb E}[ {{\bf{x}}{{\bf{x}}^H}} ] = {\bf{I}}$ is the transmitted signal from the Tx, $s \le \min \{ {{K_{\rm{T}}},{K_{\rm{R}}}} \}$ is the number of data streams, and ${{\bf{F}}_{{\rm{BB}}}} \in {\mathbb C}{^{{K_{\rm{T}}} \times s}}$ (${{\bf{W}}_{{\rm{BB}}}} \in {\mathbb C}{^{{K_{\rm{R}}} \times s}}$) and ${{\bf{F}}_{{\rm{RF}}}} \in {\mathbb C}{^{{N_{\rm{T}}} \times {K_{\rm{T}}}}}$ (${{\bf{W}}_{{\rm{RF}}}} \in {\mathbb C}{^{{N_{\rm{R}}} \times {K_{\rm{R}}}}}$) are the digital and analog beamforming at the Tx (Rx), respectively. ${\bf H}\in {\mathbb C}^{{N_{\rm R}}\times{N_{\rm T}}}$ is the composite channel between the Tx and Rx, and ${\bf n}\in {\mathbb C}^{{N_{\rm R}}\times 1}\sim {\mathcal{CN}}(0,\sigma^2 {\bf I})$ is the additive noise. The power constraint at the Tx is $\| {{{\bf{F}}_{{\rm{RF}}}}{{\bf{F}}_{{\rm{BB}}}}{\bf{x}}} \|_F^2 \le E$, where $E$ is the total transmit power.

\subsection{Channel Model}\label{sec-2.1}

\textcolor{black}{In this study, we adopt the simplified Saleh-Valenzuela model \cite{channel-model}, \cite{channel-model-Rev}, which incorporates the Rician factor to describe the mmWave channel.} As shown in Fig. \ref{Fig.system_model}, the composite channel ${\bf H}$ contains the direct channel ${\bf H}_{\rm D}$ and the segmented reflection channels, ${\bf H}_{{\rm T},k}$ and ${\bf H}_{{\rm R},k}$. The Tx--RIS $k$ channel can be expressed as
\begin{equation}\label{eq-H_Tk}
\begin{aligned}
  {{\bf{H}}_{{\rm{T}},k}} = \sqrt {{N_{\rm{T}}}{N_{{\rm{S,}}k}}} {g_{{\rm{T}},k}}&\left( {{I_{{\rm{T}},k}}\sqrt {\frac{{{\kappa _{{\rm{T}},k}}}}{{{\kappa _{{\rm{T}},k}} + 1}}} {\bf{H}}_{{\rm{T}},k}^{{\rm{LoS}}}} \right.\\
  &\left. { + \sqrt {\frac{1}{{{\kappa _{{\rm{T}},k}} + 1}}} {\bf{H}}_{{\rm{T}},k}^{{\rm{NLoS}}}} \right),
\end{aligned}
\end{equation}
where \textcolor{black}{${g_{{\rm{T}},k}}=\frac{\lambda}{4\pi r_{{\rm T},k}}$  is the path loss\footnote{\textcolor{black}{This model results in a cascaded path loss, which is proportional to the product of distances from RIS $k$ to the Tx and the Rx for the Tx-RIS $k$-Rx channel. This result is consistent with the models proposed in \cite{W. Tang_2} and \cite{Emil} for RISs. Multiple reflections by multiple RISs incur severe cascaded path loss, particularly in mmWave systems, due to this product rule for cascaded path loss. Therefore, only single reflection by RIS is considered in this work.} } determined by the Tx-RIS $k$ distance ${r_{{\rm T},k}}$.} The Rician factor ${\kappa _{{\rm{T}},k}}$ denotes the power ratio of the LoS component ${\bf{H}}_{{\rm{T}},k}^{{\rm{LoS}}}$ to the non-LoS (NLoS) component ${\bf{H}}_{{\rm{T}},k}^{{\rm{NLoS}}}$. \textcolor{black}{${I_{{\rm{T}},k}} \in ({0,1} ]$ indicates the long-term link quality of the LoS path.} ${\bf{H}}_{{\rm{T}},k}^{{\rm{LoS}}}={{\bf{a}}_{{\rm{S,}}k}}( {\Phi _{{\rm{T}},k}^{\rm{A}}}, {\Theta _{{\rm{T}},k}^{\rm{A}}} ){\bf{a}}_{\rm{T}}^H( {\Theta _{{\rm{T}},k}^{\rm{D}}} )$ and ${\bf{H}}_{{\rm{T}},k}^{{\rm{NLoS}}} = \frac{1}{\sqrt{L_{{\rm{T}},k}}}\sum\nolimits_{l = 1}^{L_{{\rm{T}},k}} {{\beta _{{\rm{T}},k,l}}{{\bf{a}}_{{\rm{S}},k}}( {\Phi _{{\rm{T}},k,l}^{\rm{A}}},{\Theta _{{\rm{T}},k,l}^{\rm{A}}} ){\bf{a}}_{\rm{T}}^H( {\Theta _{{\rm{T}},k,l}^{\rm{D}}} )} $, where ${\beta _{{\rm{T}},k,l}}\in {\mathcal {CN}(0,1)}$, ${L_{{\rm{T}},k}}$ is the number of NLoS paths. ${\bf a}_{{\rm S},k}(\cdot)$ and ${\bf a}_{{\rm T}}(\cdot)$ are the array response vectors at RIS $k$ and the Tx, respectively. In the array response vector, ${\Phi _{{\rm{T}},k}^{\rm{A}}}=2\pi d_{\rm S}\cos {\phi _{{\rm{T}},k}^{\rm{A}}}/\lambda$, ${\Theta _{{\rm{T}},k}^{\rm{A}}}=2\pi d_{\rm S}\sin {\phi _{{\rm{T}},k}^{\rm{A}}}\sin {\theta _{{\rm{T}},k}^{\rm{A}}}/\lambda$, and ${\Theta _{{\rm{T}},k}^{\rm{D}}}=2\pi d_{\rm T}\sin {\theta _{{\rm{T}},k}^{\rm{D}}}/\lambda$, where ${\phi _{{\rm{T}},k}^{\rm{A}}}$, ${\theta _{{\rm{T}},k}^{\rm{A}}}$, and ${\theta _{{\rm{T}},k}^{\rm{D}}}$ are the vertical AoA, horizontal AoA, and AoD in the LoS channel, respectively.\par
In analogy with the Tx--RIS $k$ channel, the Rx--RIS $k$ and Tx--Rx channels can be respectively given by
\begin{equation}\label{eq-H_Rk}
\begin{aligned}
  {{\bf{H}}_{{\rm{R}},k}} = \sqrt {{N_{\rm{R}}}{N_{{\rm{S,}}k}}} {g_{{\rm{R}},k}}&\left( {I_{{\rm{R}},k}}\sqrt {\frac{{{\kappa _{{\rm{R}},k}}}}{{{\kappa _{{\rm{R}},k}} + 1}}} {\bf{H}}_{{\rm{R}},k}^{{\rm{LoS}}} \right.\\
&\left.+ \sqrt {\frac{1}{{{\kappa _{{\rm{R}},k}} + 1}}} {\bf{H}}_{{\rm{R}},k}^{{\rm{NLoS}}} \right)
\end{aligned}
\end{equation}
and
\begin{equation}\label{eq-H_D}
  {{\bf{H}}_{\rm{D}}} = \sqrt {{N_{\rm{T}}}{N_{\rm{R}}}} {g_{\rm{0}}}\left( {{I_{\rm{D}}}\sqrt {\frac{{{\kappa _{\rm{D}}}}}{{{\kappa _{\rm{D}}} + 1}}} {\bf{H}}_{\rm{D}}^{{\rm{LoS}}}+ \sqrt {\frac{1}{{{\kappa _{\rm{D}}} + 1}}} {\bf{H}}_{\rm{D}}^{{\rm{NLoS}}}} \right),
\end{equation}
where \textcolor{black}{${\bf{H}}_{{\rm{R}},k}^{{\rm{LoS}}}={{\bf{a}}_{\rm{R}}}( {\Theta _{{\rm{R}},k}^{\rm{A}}} ){\bf{a}}_{{\rm{S,}}k}^H( {\Phi _{{\rm{R}},k}^{\rm{D}}},{\Theta _{{\rm{R}},k}^{\rm{D}}} )$ and ${\bf{H}}_{{\rm{R}},k}^{{\rm{NLoS}}}$ are the LoS and NLoS components of ${\bf{H}}_{{\rm{R}},k}$, respectively; ${\bf{H}}_{\rm{D}}^{{\rm{LoS}}}={{\bf{a}}_{\rm{R}}}( {\Theta _{{\rm{R}},0}^{\rm{A}}} ){\bf{a}}_{\rm{T}}^H( {\Theta _{{\rm{T}},0}^{\rm{D}}} )$ and ${\bf{H}}_{\rm{D}}^{{\rm{NLoS}}}$ are the LoS and NLoS components of ${\bf{H}}_{\rm{D}}^{{\rm{NLoS}}}$, respectively; and ${\bf{H}}_{{\rm{R}},k}^{{\rm{NLoS}}}$ and ${\bf{H}}_{\rm{D}}^{{\rm{NLoS}}}$ have an expression similar to ${\bf{H}}_{{\rm{T}},k}^{{\rm{NLoS}}}$.} The array response for UPA can be decomposed into that of ULA as ${\bf{a}}_{{\rm{S,}}k}( \Phi ,\Theta )={\bf{a}}_{{\rm{S,v,}}k}( \Phi  )\otimes {\bf{a}}_{{\rm{S,h,}}k}( \Theta )$. \textcolor{black}{This study considers the far-field scenario. Thus, }the array response vector of ULA can be unified by
\begin{equation}\label{eq-array_res_pon}
  {\bf a}_{X}(Y) = \frac{1}{\sqrt{N_{X}}}{\left[1, e^{jY},\cdots,e^{j(N_{X}-1)Y}\right]^T},
\end{equation}
where $X\in \{\{{\rm T}\},\{{\rm R}\},\{{\rm S,v},k\},\{{\rm S,h},k\}\}$.\par


\textcolor{black}{A pilot transmission method for the channel estimation stage was proposed in \cite{channel_est} to enable the receiver to separate signals arriving from different RISs and from the uncontrolled multipath. This method facilitates individual channel estimation for ${{\bf{H}}_{{\rm{T}},k}}$, ${{\bf{H}}_{{\rm{R}},k}}$, and ${{\bf{H}}_{\rm{D}}}$ because the parameters of each path can be estimated accurately \textcolor{black}{utilizing appropriate parameter extraction algorithms like atomic norm minimization \cite{channel_est_1}. Because channel estimation is not the focus of this study, we assume the perfect channel parameters can be obtained\footnote{\textcolor{black}{In practice, only the imperfect channel parameters can be obtained because of the channel estimation error. These imperfect channel parameters will incur negative effects, which are similar to that of quantization error of RIS reflection phases and imperfect orthogonality of the RIS segmentation in the following sections. Thus, the proposal of this study is valid with imperfect channel parameters. To better elaborate the main contributions of this study, the channel estimation error is not considered.}}.}} After channel estimation, the composite channel consisting of $K$ RISs can be expressed as
\begin{equation}\label{eq-H}
  {\bf{H}} = {{\bf{H}}_{\rm{D}}} + \sum\limits_{k \in {\mathcal K}} {{{\bf{H}}_{{\rm{R}},k}}{{\bf{\Gamma }}_k}{{\bf{H}}_{{\rm{T}},k}}},
\end{equation}
where ${{\bf{\Gamma }}_k}  \in {\mathbb C}{^{{N_{{\rm{S,}}k}} \times {N_{{\rm{S,}}k}}}}$ is the reflection matrix of RIS $k$. When column-level control is used, elements in the same column of the RIS are fed with the same control circuit. Thus, ${{\bf{\Gamma }}_k} = {\bf I}\otimes {\rm{diag}}( {{e^{j{\varpi _{k,1}}}}, \cdots ,{e^{j{\varpi _{k,{N_{{\rm{S,h,}}k}}}}}}} )$, with ${{{\varpi _{k,n}}}}$ being the reflection phase of the $n$-th column. Assuming that the reflection phase is quantized by $b$ bits, we obtain ${\varpi _{k,n}}\in \{2\pi i/2^b:i=1,2,\cdots,2^b\}$. \textcolor{black}{In typical strong LoS mmWave communication systems, the LoS channel component always exists in the Tx--RIS, Rx--RIS, and Tx--Rx channels, and the Rician factor is sufficiently large, thus being able to ignore the NLoS component in such scenarios\footnote{ The NLoS channel component can be seem as a disturbance for our proposal and their effects will be evaluated in the numerical results.}}. For example, on the basis on the channel measurement in \cite{Rician-1}, the Rician factor can be set to $13.5$ dB \cite{App-mmWave-2}. From these observations, we overlook the NLoS channel by setting ${\kappa_{{\rm{T}},k}}, {\kappa_{{\rm{R}},k}}, {\kappa_{\rm{D}}}\to \infty$ in the following theoretical analysis to gain insights on channel customization and joint Tx-RISs-Rx design.\par

In conventional mmWave communication systems without RISs, the rank of the channel between Tx and Rx will be reduced to 1 and cannot increase with the number of antennas when the strong LoS component dominates the channel. In these scenarios, the rank-deficient channel cannot provide spatial multiplexing to realize multi-stream transmission. Fortunately, the introduction of RISs produces a controllable scatters channel environment. By proactively designing the reflection matrices of RISs, a favorable communication environment can be established for multi-stream transmission. Given this consideration, we first investigate the design of RISs in the following sections to customize an arbitrary rank channel in strong LoS mmWave systems. After that, with a fixed number of data streams, we jointly design the transmit power at the Tx, the digital and analog beamforming at the Tx and Rx, as well as the reflection matrix at the RISs to improve the downlink SE.

\section{Channel Customization}\label{sec-3}
For the spatial multiplexing promotion and the realization of multi-stream transmission in the strong LoS mmWave coverage, the composite channel should be devised so that channel rank is no less than the data stream number. By applying the SVD, the composite channel can be represented by
 \begin{equation}\label{eq-SVD}
   {\bf{H}} = {\hat{\bf U}}{\rm{blkdiag}}\left\{ {{\bf{\Lambda }},{{\bf{0}}_{\left( {{N_{\rm{T}}} - Q} \right) \times \left( {{N_{\rm{R}}} - Q} \right)}}} \right\} {\hat{\bf V}}^H  = {\bf{U\Lambda }}{{\bf{V}}^H},
\end{equation}
where ${\bf{\Lambda }} = {\rm{diag}}( {\sqrt {{\lambda _1}} , \cdots ,\sqrt {{\lambda _Q}} } )$ is a $Q \times Q$ diagonal matrix with $\sqrt {{\lambda _n}}$ being the $n$-th largest non-zero singular value, ${\bf U}$ and ${\bf V}$ are formed by the first $Q$ columns of the unitary matrices ${\hat{\bf U}}\in {\mathbb C}^{{N_{T}}\times{N_{\rm T}}}$ and ${\hat{\bf V}}\in {\mathbb C}^{{N_{R}}\times{N_{\rm R}}}$, respectively. According to the definition of rank, the number of non-zero singular values, $Q$, is the rank of ${\bf H}$. Consequently, even a tiny singular value can contribute one rank to the channel. In practical communication systems, the link quality of the sub-channel mode is proportional to its corresponding singular value. Thus, the sub-channel mode with a tiny singular value must not be used for transmitting data because of the poor channel condition. In other words, channel rank cannot fully represent the maximal number of data streams. To enhance the characterization of the composite channel ${\bf{H}}$, we adopt the effective rank \cite{Erank}, which is defined by
\begin{equation}\label{eq-Erank}
  {\rm{Erank}}\left( {\bf{H}} \right) = \exp \left( { - \sum\limits_n {{{\bar \lambda }_n}\ln {{\bar \lambda }_n}} } \right),
\end{equation}
where ${\bar \lambda _n} = \sqrt {{\lambda _n}} /\sum\nolimits_m {\sqrt {{\lambda _m}} } $ is the singular value distribution of ${\bf H}$. With the effective rank, the contribution of a tiny singular value can be eliminated given that $\mathop {\lim }\limits_{{{\bar \lambda }_n} \to 0} {{\bar \lambda }_n}\ln {{\bar \lambda }_n} = 0$. As proven in \cite{Erank}, the effective rank for a rank $Q$ matrix will be maximal and equal to $Q$ if and only if all non-zero singular values are equal. On the basis of the relation between rank and effective rank given an arbitrary number of data streams $s$, we define the truncated condition number as
\begin{equation}\label{eq-trun-cond-numb}
  T_{s} = \frac{\sqrt{\lambda_1}}{\sqrt{\lambda_s}}.
\end{equation}
When $\lambda_i=\lambda_1$ for $\forall i\leq s$ and $\lambda_j=0$ for $\forall j>s$, the truncated condition number achieves its minimum, that is, 1, and the effective rank is equal to the rank. In this case, the composite channel is \textcolor{black}{fair} for the $s$-stream transmission. Therefore, we aim to minimize the truncated condition number by optimizing the deployment location and reflection matrices of RISs. To achieve this goal, the minimal number of elements for each RIS will also be derived for combating the path loss.\par

\textcolor{black}{In the strong LoS mmWave MIMO systems where ${\kappa_{{\rm{T}},k}}, {\kappa_{{\rm{R}},k}}, {\kappa_{\rm{D}}}\to \infty$, the composite channel in \eqref{eq-H} can be simplified as
\begin{equation}\label{eq-H-S1}
  \begin{aligned}
{\bf{H}} &= {I_{\rm{D}}}\sqrt {{N_{\rm{T}}}{N_{\rm{R}}}} {g_{\rm{0}}}{{\bf{a}}_{\rm{R}}}\left( {\Theta _{{\rm{R}},0}^{\rm{A}}} \right){\bf{a}}_{\rm{T}}^H\left( {\Theta _{{\rm{T}},0}^{\rm{D}}} \right)\\
 &+ \sum\limits_{k \in {\mathcal K}} {I_{{\rm{T}},k}} {I_{{\rm{R}},k}}\sqrt {{N_{\rm{R}}}{N_{\rm{T}}}} {N_{{\rm{S}},k}}{g_{{\rm{R}},k}}g_{{\rm{T}},k}^*{{\bf{a}}_{\rm{R}}}\left( {\Theta _{{\rm{R}},k}^{\rm{A}}} \right)\\
 &\times {\bf{a}}_{{\rm{S}},k}^H\left( {\Phi _{{\rm{R}},k}^{\rm{D}}},{\Theta _{{\rm{R}},k}^{\rm{D}}} \right){{\bf{\Gamma }}_k}{{\bf{a}}_{{\rm{S}},k}}\left( {\Phi _{{\rm{T}},k}^{\rm{A}}},{\Theta _{{\rm{T}},k}^{\rm{A}}} \right){\bf{a}}_{\rm{T}}^H\left( {\Theta _{{\rm{T}},k}^{\rm{D}}} \right).
\end{aligned}
\end{equation}
Let $f( {{{\bf{\Gamma }}_k}} ) \buildrel \Delta \over = {{N_{{\rm{S}},k}}{\bf{a}}_{{\rm{S}},k}^H( {\Phi _{{\rm{R}},k}^{\rm{D}}},{\Theta _{{\rm{R}},k}^{\rm{D}}} ){{\bf{\Gamma }}_k}{{\bf{a}}_{{\rm{S}},k}}({\Phi _{{\rm{T}},k}^{\rm{A}}}, {\Theta _{{\rm{T}},k}^{\rm{A}}} )}$, ${\alpha _k} \buildrel \Delta \over = {I_{{\rm{T}},k}} {I_{{\rm{R}},k}}\sqrt {{N_{\rm{T}}}{N_{\rm{R}}}} {g_{{\rm{R}},k}}{g^*_{{\rm{T}},k}}f( {{{\bf{\Gamma }}_k}} )$, and ${\alpha _0} \buildrel \Delta \over = {I_{\rm{D}}}\sqrt {{N_{\rm{T}}}{N_{\rm{R}}}} {g_0}$ being the array gain of RIS $k$, the reconfigurable cascaded path gain of the Tx--RIS $k$--Rx channel, and the path gain of the direct Tx--Rx channel,} respectively, \eqref{eq-H-S1} can be unified as
\begin{equation}\label{eq-H-LoS-dom}
  {\bf{H}} = \sum\limits_{k = 0}^K {{\alpha _k}{{\bf{a}}_{\rm{R}}}\left( {\Theta _{{\rm{R}},k}^{\rm{A}}} \right){\bf{a}}_{\rm{T}}^H\left( {\Theta _{{\rm{T}},k}^{\rm{D}}} \right)}  = {{\bf{A}}_{\rm{R}}}{\bf{\Sigma A}}_{\rm{T}}^H,
\end{equation}
where ${{\bf{A}}_{\rm{R}}} = [ {{{\bf{a}}_{\rm{R}}}( {\Theta _{{\rm{R}},0}^{\rm{A}}} ),{{\bf{a}}_{\rm{R}}}( {\Theta _{{\rm{R}},1}^{\rm{A}}} ), \cdots ,{{\bf{a}}_{\rm{R}}}( {\Theta _{{\rm{R}},K}^{\rm{A}}} )} ]$ is the array response matrix determined by the AoAs at the Rx, ${{\bf{A}}_{\rm{T}}} = [ {{e^{\angle {\alpha _{\rm{0}}}}}{{\bf{a}}_{\rm{T}}}( {\Theta _{{\rm{T}},0}^{\rm{D}}} ),{e^{\angle {\alpha _{\rm{1}}}}}{{\bf{a}}_{\rm{T}}}( {\Theta _{{\rm{T}},1}^{\rm{D}}} ), \cdots ,{e^{\angle {\alpha _K}}}{{\bf{a}}_{\rm{T}}}( {\Theta _{{\rm{T}},K}^{\rm{D}}} )} ]$ is the phase-shifted array response matrix determined by the AoDs at the Tx, and ${\bf{\Sigma }} = {\rm{diag}}( { {| {{\alpha _{\rm{0}}}} |,| {{\alpha _1}} |, \cdots ,| {{\alpha _K}} |} } )$. The simplified composite channel in \eqref{eq-H-LoS-dom} is similar to the SVD in \eqref{eq-SVD}. When the columns of ${{\bf{A}}_{\rm{R}}}$ (${{\bf{A}}_{\rm{T}}}$) that correspond to the non-zero diagonal elements in ${\bf \Sigma}$ are pair-wisely orthogonal, ${{\bf{A}}_{\rm{R}}}{\bf{\Sigma A}}_{\rm{T}}^H$ happens to be the SVD of ${\bf H}$, and the composite channel can be decoupled as several interference-free sub-channels. From this observation, we will investigate the orthogonality among columns in ${{\bf{A}}_{\rm{R}}}$ and in ${{\bf{A}}_{\rm{T}}}$ and customize the composite channel in terms of singular values for arbitrary multi-stream transmission. \textcolor{black}{Given that the number of antennas affects the orthogonality of the array response vectors in ${{\bf{A}}_{\rm{R}}}$ and ${{\bf{A}}_{\rm{T}}}$, we consider three scenarios with different antenna configurations:  massive MIMO (M-MIMO), unilateral M-MIMO, and conventional MIMO.}

\subsection{Channel Customization in \textcolor{black}{M-MIMO}}\label{sec-3.1}
\textcolor{black}{In this study, M-MIMO refers to the scenario wherein the number of antennas at the Tx and Rx is sufficiently large, while the number of RISs is limited, that is, $K \ll {N_{\rm{R}}},{N_{\rm{T}}} $.}\par


For clarity of presentation, we first consider the general case where the Tx, RIS $k$, and Rx are not collinear\footnote{\textcolor{black}{When the array resolution of the $N_{\rm T}$-antenna Tx is defined as $\frac{\pi }{{{N_{\rm{T}}}}}$, the Tx, RIS $k$, and the Rx are regarded as collinear when the difference between the trigonometric LoS AoDs of the Tx--Rx and Tx--RIS $ k$ channels is less than the array resolution of Tx, that is, $| {\Theta _{{\rm{T}},0}^{\rm{D}} - \Theta _{{\rm{T}}, k}^{\rm{D}}} | \le \frac{\pi }{{{N_{\rm{T}}}}}$.}}. When the Tx and Rx are equipped with an extremely large number of antennas, we can easily prove that the array response vectors are asymptotically orthogonal \cite{bit_par}, that is,
\begin{equation}\label{eq-Asym_ortho}
  \left\{ \begin{aligned}
&{\bf{a}}_{\rm{T}}^H\left( {\Theta _{{\rm{T}},n}^{\rm{D}}} \right){{\bf{a}}_{\rm{T}}}\left( {\Theta _{{\rm{T}},m}^{\rm{D}}} \right) \to 0, n \ne m\\
&{\bf{a}}_{\rm{R}}^H\left( {\Theta _{{\rm{R}},i}^{\rm{A}}} \right){{\bf{a}}_{\rm{R}}}\left( {\Theta _{{\rm{R}},j}^{\rm{A}}} \right) \to 0, i \ne j
\end{aligned} \right..
\end{equation}
The asymptotical orthogonality of array response vectors at the Tx and Rx means that \textcolor{black}{the inter-beam inference at the Tx and Rx can be neglected, and} \eqref{eq-H-LoS-dom} is the SVD of $\bf H$. Hence, $\{ {| {{\alpha _k}} |} \}_{k = 0}^K$ are the singular values and $K+1$ is the maximal rank that can be achieved. Considering that $\alpha_0$ is the constant path gain of the direct Tx--Rx channel that cannot be adjusted, the cascaded path gain of the Tx--RIS $k$--Rx channel should be carefully tuned to customize a favorable channel that has the maximum effective rank so that $| {{\alpha _k}} | = | {{\alpha _0}} |$ for $k \in {\mathcal K}$. Unfolding $| {{\alpha _k}} |$ and $| {{\alpha _0}} |$, the condition $| {{\alpha _k}} | = | {{\alpha _0}} |$ can be rewritten as
\textcolor{black}{\begin{equation}\label{eq-f_equal_to}
  \left| {f\left( {{{\bf{\Gamma }}_k}} \right)} \right| = \frac{{I_{{\rm{D}}}} {{g_0}}}{{I_{{\rm{T}},k}} {I_{{\rm{R}},k}}{{g_{{\rm{R}},k}}{g_{{\rm{T}},k}}}} =  {4\pi } \frac{{{r_{{\rm{R}},k}}{r_{{\rm{T}},k}}}}{{I_k} {{r_0}}\lambda},
\end{equation}
where the second equation is derived with the relation between path loss $g$ and distance $r$, and $I_k={I_{{\rm{T}},k}} {I_{{\rm{R}},k}}/{I_{{\rm{D}}}}$.} The absolute array gain of the RIS, $| {f( {{{\bf{\Gamma }}_k}} )} |$, is adjustable by the reflection matrix ${{{\bf{\Gamma }}_k}}$. \textcolor{black}{Therefore, the prerequirement is that the maximum of $| {f( {{{\bf{\Gamma }}_k}} )} |$ should be no less than $ {4\pi } \frac{{{r_{{\rm{R}},k}}{r_{{\rm{T}},k}}}}{I_k{{r_0}}\lambda}$ to make the condition \eqref{eq-f_equal_to} hold.} The maximal achievable array gain of the RIS is provided in \emph{Lemma} \ref{lemma-1}.
\begin{lemma}\label{lemma-1}
\textcolor{black}{For a sufficient large $N_{{\rm S},k}$-element RIS with column-level control, when the Tx, RIS, and Rx are at the same altitude, the maximum value of the array gain, that is, ${f_{\max}\left( {{{\bm{\Gamma }}_k}} \right)} = {N_{{\rm{S}},k}}$, can be achieved by setting the reflection phase of the elements in column $n$ as $\varpi _{k,n}^{{\rm{c,opt}}} = ( {n - 1} )( {\Theta _{{\rm{R}},k}^{\rm{D}} - \Theta _{{\rm{T}},k}^{\rm{A}}} )$, $\forall n$. If the reflection phase is quantized as $\varpi _{k,n}^{{\rm{d,opt}}}$ by $b$ bits, that is
\begin{equation}\label{eq-optimal_phase_d_L}
		- \frac{\pi }{{{2^b}}} \le \varpi _{k,n}^{{\rm{d,opt}}} - \varpi _{k,n}^{{\rm{c,opt}}} \le \frac{\pi }{{{2^b}}},
\end{equation}
then the maximum value of the array gain is approximated by
\begin{equation}\label{eq-f_max}
   {f_{\max}\left( {{{\bf{\Gamma }}_k}} \right)} \approx {N_{{\rm{S}},k}}{\rm sinc} (\frac{\pi }{{{2^b}}}),
\end{equation}
where ${\rm sinc} (\frac{\pi }{{{2^b}}}) = \sin(\frac{\pi }{{{2^b}}})/\frac{\pi }{{{2^b}}}$.}
\end{lemma}
\begin{IEEEproof}
\textcolor{black}{See Appendix \ref{App:A}.}
\end{IEEEproof}\par
\emph{Lemma} \ref{lemma-1} shows that the maximum of $| {f( {{{\bf{\Gamma }}_k}} )} |$ is determined by the scale of RISs, ${N_{{\rm{S}},k}}$, and the quantization bits $b$. When $b \to \infty$, ${f_{\max}\left( {{{\bf{\Gamma }}_k}} \right)} \to {N_{{\rm{S}},k}}$ can be derived for \eqref{eq-f_max}, which is consistent with the maximum array gain achieved with the continuously controlled reflection phase. Now, given the fixed quantization bits, we analyze the minimum scale of RISs that is required for \eqref{eq-f_equal_to} to combat the path loss and quantization loss. When the Tx and RISs are settled, $\{ {{r_{{\rm{T}},k}}} \}_{k = 1}^K$ are constants while ${r_0}$ and $\{ {{r_{{\rm{T}},k}}} \}_{k = 1}^K$ depend on the position of the Rx ${{\bf{e}}_{\rm{R}}}\in {\mathcal A}$. To ensure that the favorable channel can be established for the Rx in every position of the coverage area, according to \eqref{eq-f_equal_to}, the number of elements of RIS $k$ should satisfy
\textcolor{black}{
\begin{equation}\label{eq-N_s,k_cons}
  {N_{{\rm{S}},k}}{\rm sinc} (\frac{\pi }{{{2^b}}})\ge \mathop {\max }\limits_{{{\bf{e}}_{\rm{R}}}\in {\mathcal A}} \left( { {4\pi } \frac{{{r_{{\rm{R}},k}}{r_{{\rm{T}},k}}}}{I_k{{r_0}}\lambda}} \right).
\end{equation}}
The right term of \eqref{eq-N_s,k_cons} varies in different scenarios because it depends on the geometric position and shape of the coverage area ${\mathcal A}$. To eliminate the tedious calculation for the rigorous lower bound constraint of ${N_{{\rm{S}},k}}$ and obtain an easy system implementation, we set the number of elements of RIS $k$ as
\textcolor{black}{\begin{equation}\label{eq-N_s,k_equal}
  {N_{{\rm{S}},k}} = \left\lceil {\frac{{ {4{\pi}} }}{{\rm sinc} (\frac{\pi }{{{2^b}}})}\frac{{r_{{\rm{R}},k}^{\max }{r_{{\rm{T}},k}}}}{I_k{r_0^{\min }}\lambda}} \right\rceil,
\end{equation}}
where ${{r_0^{\min }}}=\mathop {\min }\limits_{{{\bf{e}}_{\rm{R}}}\in {\mathcal A}} ( r_0 )$ and $r_{{\rm{R}},k}^{\max }=\mathop {\max }\limits_{{{\bf{e}}_{\rm{R}}}\in {\mathcal A}} ( r_{{\rm T},k} )$.  \textcolor{black}{A special case for ${N_{{\rm{S}},k}}$ can be found in \cite[ Corollary 4]{number} when the positions of the Tx and Rx are fixed and the discrete phase shifter of the RIS is not considered.} Notably, ${N_{{\rm{S}},k}}$ in \eqref{eq-N_s,k_equal} is a surplus guarantee for condition \eqref{eq-N_s,k_cons}. When the reflection phase of the RIS is properly designed, such ${N_{{\rm{S}},k}}$ enables Rx to achieve $| {{\alpha _k}} | = | {{\alpha _0}} |$ anywhere in the coverage because the array gain loss results from limited reflection phase quantization and the extra path loss brought by the distance product in the reflection link is compensated by more array gains provided by the increment of RIS elements.\par

On the basis of the RIS scale requirement \eqref{eq-N_s,k_equal}, we now focus on the reflection phase design to customize a channel that has the maximal effective rank, which means that \eqref{eq-f_equal_to} should be satisfied for $k \in {\mathcal K}$. We split $ {f\left( {{{\bf{\Gamma }}_k}} \right)} $ into two parts as
\begin{equation}\label{eq-f_two_parts}
\begin{aligned}
   {f\left( {{{\bf{\Gamma }}_k}} \right)}&={{N_{{\rm{S,v,}}k}}}\sum\limits_{n = 1}^{{\gamma _k}} {{e^{j\left( {{\varpi _{k,n}} - \left( {n - 1} \right)\left( {\Theta _{{\rm{R}},k}^{\rm{D}} - \Theta _{{\rm{T}},k}^{\rm{A}}} \right)} \right)}}}\\
   &+{{N_{{\rm{S,v,}}k}}}\sum\limits_{n = {\gamma _k} + 1}^{{N_{{\rm{S,h}},k}}} {{e^{j\left( {{\varpi _{k,n}} - \left( {n - 1} \right)\left( {\Theta _{{\rm{R}},k}^{\rm{D}} - \Theta _{{\rm{T}},k}^{\rm{A}}} \right)} \right)}}},
\end{aligned}
\end{equation}
\textcolor{black}{where ${\gamma _k} = \lceil {\frac{{ {4{\pi }} }}{{\rm sinc} (\frac{\pi }{{{2^b}}})}\frac{{{r_{{\rm{R}},k}}{r_{{\rm{T}},k}}}}{I_k{{r_0}}\lambda{{N_{{\rm{S,v,}}k}}}}} \rceil$ is chosen so that the first part can be approximated as $ {4\pi } \frac{{{r_{{\rm{R}},k}}{r_{{\rm{T}},k}}}}{I_k{{r_0}}\lambda}$} when the reflection phase of the first ${\gamma _k}$ columns of RIS $k$ is designed as
\begin{equation}\label{eq-first_gamma_k}
  {\varpi _{k,n}} = \varpi _{k,n}^{{\rm{d,opt}}}.
\end{equation}
To achieve \eqref{eq-f_equal_to}, the second term in the right of \eqref{eq-f_two_parts} should be equal to 0. An infinite number of solutions exist for the design of the reflection phase of the last ${N_{{\rm{S,h}},k}} - {\gamma _k}$ RIS columns. \textcolor{black}{Among these solutions, the most economical approach is to keep these columns undesigned, that is, ${\varpi _{k,n}}=0$ for $n\in \{{\gamma _k}+1,\cdots,N_{{\rm S,h},k}\}$.} When ${N_{{\rm{S,h}},k}}$ is sufficiently large and ${\gamma _k}$ is limited, according to the asymptotical orthogonality of array response vectors, we have
\begin{equation}\label{eq:turn-off}
  \sum\limits_{n = {\gamma _k} + 1}^{{N_{{\rm{S,h}},k}}} {{e^{j\left( {{\varpi _{k,n}} - \left( {n - 1} \right)\left( {\Theta _{{\rm{R}},k}^{\rm{D}} - \Theta _{{\rm{T}},k}^{\rm{A}}} \right)} \right)}}} \to 0.
\end{equation}
\textcolor{black}{At this point, the composite channel $\bf H$ can be customized to have the maximum effective rank and equal singular values by utilizing \eqref{eq-first_gamma_k} to design the reflection phases of the first ${\gamma _k}$ columns of RIS $k$.}\par

More than the aforementioned customization strategy, the composite channel can be flexibly shaped to let the effective rank be $s$ when the Tx transmits $s$ ($1 \le s \le K + 1$) data streams. \textcolor{black}{Specifically, following the ascending order of cascaded path losses, we divide the RISs into two sets, ${{\mathcal K}_ \bot }$ and ${\bar{\mathcal K}_ \bot }={{\mathcal K} }\backslash {{\mathcal K}_ \bot }$, where $| {{{\mathcal K}_ \bot }} | = s - 1$ and ${\bar{\mathcal K}_ \bot }$ contains RISs with more severe path loss.} For $i \in {{\mathcal K}_ \bot }$, we design the reflection phases of RIS $i$ according to \eqref{eq-first_gamma_k} to let $| {{\alpha _i}} | = | {{\alpha _0}} |$. \textcolor{black}{For $i \in {\bar{\mathcal K}_ \bot }$, RIS $i$ is inactivated so that $| {{\alpha _i}} | \to 0$ can be achieved according to the asymptotical orthogonality of array response vectors.}\par

So far, channel customization has been discussed for non-collinear Tx, RIS $k$, and Rx. When the difference between the LoS AoDs of the Tx--Rx and Tx--RIS $\bar k$ channel is less than the array resolution of Tx, that is, $| {\Theta _{{\rm{T}},0}^{\rm{D}} - \Theta _{{\rm{T}},\bar k}^{\rm{D}}} | \le \frac{\pi }{{{N_{\rm{T}}}}}$, the Tx, RIS $\bar k$, and Rx are regarded as collinear. In this case, ${{\bf{a}}_{\rm{T}}}( {\Theta _{{\rm{T}},0}^{\rm{D}}} )$ and ${{\bf{a}}_{\rm{T}}}( {\Theta _{{\rm{T}},\bar k}^{\rm{D}}} )$ are highly correlated, and the maximum rank of $\bf H$ is reduced from $K+1$ to $K$.

\subsection{Channel Customization in Unilateral \textcolor{black}{M-MIMO}}\label{sec-3.2}
\textcolor{black}{The unilateral M-MIMO refers to the scenario wherein the number of antennas at the Tx is sufficiently large. The number of RISs is limited but larger than the number of antennas at the Rx, that is, ${N_{\rm{R}}} < K \ll {N_{\rm{T}}}  $. The maximal rank in the unilateral M-MIMO is $N_{\rm R}$.}\par
In the unilateral M-MIMO scenario, the asymptotical orthogonality is still tenable for array response vectors in ${\bf A}_{\rm T}$ but invalid for that in ${\bf A}_{\rm R}$ due to the limited number of antennas at the Rx. Considering that ${N_{\rm{R}}} < K$, the number of columns in ${\bf A}_{\rm R}$ is larger than its dimension, which means that the column vectors in ${\bf A}_{\rm R}$ must be linearly dependent. Thus, \eqref{eq-H-LoS-dom} is not the SVD of the composite channel.\par

Here, our target is to customize an effective rank $s$ channel that can be explicitly expressed in the form of SVD by selecting $s-1$ RISs that generate good orthogonality among array response vectors in ${\bf A}_{\rm R}$ and reducing the array gains of the remaining RISs to avoid disturbance. Specifically, all RISs are divided into two sets, ${{\mathcal K}_ \bot }$ and ${\bar{\mathcal K}_ \bot }$, which satisfy
\begin{equation}\label{eq-two_sets_in U}
  \left\{ \begin{aligned}
&\left| {{\bf{a}}_{\rm{R}}^H\left( {\Theta _{{\rm{R}},i}^{\rm{A}}} \right){{\bf{a}}_{\rm{R}}}\left( {\Theta _{{\rm{R}},j}^{\rm{A}}} \right)} \right| \approx 0,{\rm{      }}\quad i,j \in {{\mathcal K}_ \bot } \cup \left\{ 0 \right\},i \ne j,\\
&\left| {{\alpha _k}} \right| = 0, \qquad\qquad\qquad\qquad\; k \in {\bar{\mathcal K}_ \bot },
\end{aligned} \right.
\end{equation}
where $| {{\mathcal K}_ \bot } | = s - 1$, ${{\mathcal K}_ \bot } \cup {\bar{\mathcal K}_ \bot } = {\mathcal K}$, and $ {{\mathcal K}_ \bot } \cap {\bar{\mathcal K}_ \bot } = \emptyset $. On the basis of \eqref{eq-two_sets_in U}, the composite channel can be expressed in terms of SVD, that is,
\begin{equation}\label{eq_H_SVD_in_U}
  {\bf{H}} = \sum\limits_{k \in {{\mathcal K}_ \bot } \cup \left\{ 0 \right\}} {{\alpha _k}{{\bf{a}}_{\rm{R}}}\left( {\Theta _{{\rm{R}},k}^{\rm{A}}} \right){\bf{a}}_{\rm{T}}^H\left( {\Theta _{{\rm{T}},k}^{\rm{D}}} \right)}  = {{\bf{A}}_{{\rm{R,}} \bot }}{{\bf{\Sigma }}_ \bot }{\bf{A}}_{{\rm{T,}} \bot }^H,
\end{equation}
where ${{\bf{\Sigma }}_ \bot }$ is a diagonal matrix whose diagonal elements are $\{\alpha_k, {k \in {{\mathcal K}_ \bot } \cup \{ 0 \}}\}$. ${{\bf{A}}_{{\rm{R,}} \bot }}$ and ${\bf{A}}_{{\rm{T,}} \bot }$ are composed of orthogonal array response vectors \textcolor{black}{that correspond to} $\{\alpha_k, {k \in {{\mathcal K}_ \bot } \cup \{ 0 \}}\}$. The key for channel customization \eqref{eq_H_SVD_in_U} lies in the set division that satisfies \eqref{eq-two_sets_in U}. When the columns of a matrix $\bf A$ are pair-wisely orthogonal, ${{{\bf{A}}^H}{\bf{A}} = {\bf{I}}}$. Therefore, to obtain ${{\bf{A}}_{{\rm{R,}} \bot }}$ while avoiding $| {\Theta _{{\rm{T}},0}^{\rm{D}} - \Theta _{{\rm{T}},\bar k}^{\rm{D}}} | \le \frac{\pi }{{{N_{\rm{T}}}}}$, the RISs segmentation can be formulated as\footnote{\textcolor{black}{This RIS segmentation problem always has feasible solutions because the minimum of $\| {{{\bf{A}}^H}{\bf{A}} - {\bf{I}}} \|_F^2$ can always be achieved within the limited number of possible combinations. After the RIS segmentation, ${{{\bf{A}}^H}{\bf{A}} \approx {\bf{I}}}$ not ${{{\bf{A}}^H}{\bf{A}} = {\bf{I}}}$ is obtained due to imperfect orthogonality, incurring inter-stream interference. Such inter-stream interference is inevitable, and thus, we aim to customize the channel with the possible minimum inter-stream interference. \textcolor{black}{Considering that the channel is customized for $s$-stream transmission and inter-steam interference caused by imperfect orthogonality is acceptable, the objective function is not constrained by a prescribed threshold that may make the problem infeasible.} } }
\begin{equation}\label{eq-RIS_div_orig}
  \begin{aligned}
\left\{ {{{\bf{A}}_{{\rm{R,}} \bot }},{{\mathcal K}_ \bot }} \right\} &= \mathop {\arg \min }\limits_{\left\{ {{\bf{A}},{{\mathcal K}_{s - 1}}} \right\}} \left\| {{{\bf{A}}^H}{\bf{A}} - {\bf{I}}} \right\|_F^2\\
{\rm{{s.t.}}} \qquad {\bf{A}} =& \left[ {{{\bf{a}}_{\rm{R}}}\left( {\Theta _{{\rm{R}},0}^{\rm{A}}} \right),{{\bf{a}}_{\rm{R}}}\left( {\Theta _{{\rm{R}},{k_1}}^{\rm{A}}} \right), \cdots ,{{\bf{a}}_{\rm{R}}}\left( {\Theta _{{\rm{R}},{k_{s - 1}}}^{\rm{A}}} \right)} \right],\\
{{\mathcal K}_{s - 1}} =& \left\{ {{k_i}} \right\}_{i = 1}^{s - 1} \subseteq {\mathcal K}\backslash \left\{ {\bar k} \right\}.
\end{aligned}
\end{equation}
The optimal solution for the combinatorial optimization problem \eqref{eq-RIS_div_orig} requires exhaustive search over ${\mathcal K}\backslash \{ {\bar k} \}$, which results in high search and computation complexity. To reduce this overhead, we resort to greedy search for \eqref{eq-RIS_div_orig}. The greedy search algorithm is started by selecting RIS $k_{1}$ from ${\mathcal K}\backslash \{ {\bar k} \}$ to ensure that ${{{\bf{a}}_{\rm{R}}}( {\Theta _{{\rm{R}},{k_1}}^{\rm{A}}} )}$ and ${{{\bf{a}}_{\rm{R}}}( {\Theta _{{\rm{R}},0}^{\rm{A}}} )}$ are approximately orthogonal, as discussed in \eqref{eq-two_sets_in U}. This process can be described by
\begin{equation}\label{eq-RIS_div_1}
  \begin{aligned}
\left\{ {{{\bf{A}}_1},{{\mathcal K}_1^* }} \right\} &= \mathop {\arg \min }\limits_{\left\{ {{\bf{A}},{{\mathcal K}_{ 1}}} \right\}} \left\| {{{\bf{A}}^H}{\bf{A}} - {\bf{I}}} \right\|_F^2\\
{\rm{{s.t.}}} \qquad &{\bf{A}} = \left[ {{{\bf{a}}_{\rm{R}}}\left( {\Theta _{{\rm{R}},0}^{\rm{A}}} \right),{{\bf{a}}_{\rm{R}}}\left( {\Theta _{{\rm{R}},{k_1}}^{\rm{A}}} \right)} \right],\\
&{{\mathcal K}_{ 1}} = \left\{ {{k_1}} \right\} \subseteq {\mathcal K}\backslash \left\{ {\bar k} \right\}.
\end{aligned}
\end{equation}
After $ {{\bf{A}}_1}$ and ${{\mathcal K}_1^* } $ are found, another RIS denoted by $k_{2}$ can be obtained by executing \eqref{eq-RIS_div_1} again with $ {{\bf{A}}}$ and ${{\mathcal K}_{ 1}}$ replaced by $[ {{\bf{A}}_1,{{\bf{a}}_{\rm{R}}}( {\Theta _{{\rm{R}},{k_2}}^{\rm{A}}} )} ]$ and ${{\mathcal K}_1^* }\cup \{k_2\} $, respectively. This greedy search process is performed until $s-1$ RISs, $ \{ {{k_i}} \}_{i = 1}^{s - 1}$, are selected. After that, we minimize the truncated condition number to customize the effective rank $s$ channel by letting $| {{\alpha _i}} | = | {{\alpha _0}} |$ for $i\in {\mathcal K}_\bot$ and $| {{\alpha _i}} | = 0$ for $ i \in {\bar{\mathcal K}_ \bot }$. The procedure of the greedy search-based RISs segmentation is summarized in Algorithm 1.
\begin{algorithm}[htb]
\caption{Greedy search-based RISs segmentation}
\hspace*{0.02in} {\bf Input:} AoDs at the TX $\left\{ {\Theta _{{\rm{T}},k}^{\rm{D}}} \right\}_{k = 0}^K$, AoAs at the Rx $\left\{ {\Theta _{{\rm{R}},k}^{\rm{A}}} \right\}_{k = 0}^K$, the set of RISs $\mathcal K$, the required effective rank $s$, and the number of antennas at the Tx $N_{\rm T}$.\\
\hspace*{0.02in} {\bf Output:} ${{\mathcal K}_ \bot }$ and ${\bar{\mathcal K}_ \bot }={{\mathcal K} }\backslash {{\mathcal K}_ \bot }$.
\label{alg:Framwork}
\begin{algorithmic}[1]
\State Initial: ${\mathcal K}_\bot = \emptyset$, ${\bf{A}} = \left[ {{{\bf{a}}_{\rm{R}}}\left( {\Theta _{{\rm{R}},0}^{\rm{A}}} \right)} \right]$, and $i=1$;
\State Find RIS $\bar k$ that are collinear with Tx and Rx, that is, $\left| {\Theta _{{\rm{T}},0}^{\rm{D}} - \Theta _{{\rm{T}},\bar k}^{\rm{D}}} \right| \le \frac{\pi }{{{N_{\rm{T}}}}}$;
\While{$i \le s - 1$}
\State  ${k_i} = \mathop {\arg \min }\limits_{k \in {\mathcal K}\backslash \left\{ {\bar k} \right\}} \left\| {{{\left[ {{\bf{A}},{{\bf{a}}_{\rm{R}}}\left( {\Theta _{{\rm{R}},k}^{\rm{A}}} \right)} \right]}^H}\left[ {{\bf{A}},{{\bf{a}}_{\rm{R}}}\left( {\Theta _{{\rm{R}},k}^{\rm{A}}} \right)} \right] - {\bf{I}}} \right\|_F^2$;
\State  ${{\mathcal K}_ \bot } = {{\mathcal K}_ \bot } \cup \left\{ {{k_i}} \right\}$;
\State ${\bf{A}} = \left[ {{\bf{A}},{{\bf{a}}_{\rm{R}}}\left( {\Theta _{{\rm{R}},{k_i}}^{\rm{A}}} \right)} \right]$;
\State $i=i+1$;
\EndWhile
\end{algorithmic}
\end{algorithm}

The computational complexity of the exhaustive and greedy search is compared as follows. We consider the general case where the Tx, RISs, and the Rx are not collinear. \textcolor{black}{According to Appendix \ref{App:B}, the total computational complexity of the exhaustive search-based algorithm is 
\begin{equation}\label{eq-Comp_ex_2}
	C_1={\mathcal O}\left( {\frac{{{s^2}{N_{\rm{R}}}K!}}{{\left( {s - 1} \right)!\left( {K - s + 1} \right)!}}} \right),
\end{equation}
and that of the greedy search-based algorithm is approximated by 
\begin{equation}\label{eq-Comp_greedy-3}
	C_2={\mathcal O}\left( {\frac{{{N_{\rm{R}}}\left( {4K - 3s} \right){s^3}}}{{12}}} \right).
\end{equation}}
Comparing \eqref{eq-Comp_greedy-3} with \eqref{eq-Comp_ex_2}, we can see that adopting the greedy search-based algorithm can significantly reduce the computational complexity because the factorial operation is eliminated. To obtain an intuitive sense about the overhead reduction, by assuming $s=N_{\rm R}=8$ and $K=25$, we have ${C_1}/{C_2}\approx 9500$.

\subsection{Channel Customization in Conventional MIMO}\label{sec-3.3}
The conventional MIMO refers to the scenario where the number of antennas at the Tx and Rx is limited. The number of RISs is larger than the number of antennas at the Rx but smaller than that at the Tx, that is, ${N_{\rm{R}}} < K < {N_{\rm{T}}}$.\par
In this case, the asymptotical orthogonality between array response vectors in ${\bf A}_{\rm T}$ and in ${\bf A}_{\rm R}$ is broken due to the limited number of antennas. We have shown in Section \ref{sec-3.2} that an approximate column-orthogonal matrix ${{\bf{A}}_{{\rm{R,}} \bot }}$ can be generated through the RISs segmentation. Therefore, the only question left is \textcolor{black}{how to ensure that} array response vectors for the Tx-RIS channel are orthogonal.\par

We attempt to address this task via the system deployment of RISs. In the strong LoS mmWave communication systems, except for the direct link between the Tx--Rx channel, the columns of ${{\bf{A}}_{\rm{T}}}$ are determined by the positions of the Tx and RISs. On the basis of this observation, when the Tx is settled, we install RISs at the DFT directions of the Tx so that the AoD satisfies $\Theta _{{\rm{T}},k}^{\rm{D}} \in \{ {\frac {{2\pi i}}{{{N_{\rm{T}}}}} - \pi } \}_{i = 1}^{{N_{\rm{T}}}},k \in {\mathcal K}$. In this RISs deployment, we have
\begin{equation}\label{eq-Coven_MIMO}
  {\bf{a}}_{\rm{T}}^H\left( {\Theta _{{\rm{T}},k}^{\rm{D}}} \right){{\bf{a}}_{\rm{T}}}\left( {\Theta _{{\rm{T}},m}^{\rm{D}}} \right) = 0,
\end{equation}
for $k,m \in {\mathcal K} $ and $k \ne m$. Given the location of the Tx, ${{\bf{e}}_{\rm{T}}} = {\left[ {{x_{\rm{T}}},{y_{\rm{T}}}} \right]^T}$, RIS $k$ should be placed at ${{\bf{e}}_{{\rm{S,}}k}} = {\left[ {{x_{{\rm{S,}}k}},{y_{{\rm{S,}}k}}} \right]^T}$, where
\begin{equation}\label{eq-position_RIS}
  \left\{ \begin{aligned}
&{x_{{\rm{S,}}k}} = {r_{{\rm{T,}}k}}\cos \left( {\arcsin \frac{{\Theta _{{\rm{T}},k}^{\rm{D}}}}{\pi }} \right)\\
&{y_{{\rm{S,}}k}} = {r_{{\rm{T,}}k}}\frac{{\Theta _{{\rm{T}},k}^{\rm{D}}}}{\pi }
\end{aligned} \right..
\end{equation}
So far, the composite channel in the conventional MIMO can be customized to have an effective rank $s$ and the minimal truncated condition number by RISs deployment (equation \eqref{eq-position_RIS}) and segmentation (Algorithm 1), as well as singular value modification ($\left| {{\alpha _i}} \right| = \left| {{\alpha _0}} \right|$ for $i\in {\mathcal K}_\bot$ and $\left| {{\alpha _k}} \right| = 0$ for $ k \in {\bar{\mathcal K}_ \bot }$). \textcolor{black}{Notably, this section customizes the composite channel with the objective of minimal truncated condition number to demonstrate the ability of RIS in reshaping the wireless channel. The following section applies the proposed channel customization to simplify the joint Tx-RISs-Rx design and improve the downlink SE.}

\section{{Joint Tx-RISs-Rx Design}}\label{sec-4}
Different from conventional communication systems, the introduction of RISs offers a new degree of freedom for designing transmission schemes in terms of reshaping channels. By jointly designing the hybrid beamforming at the Tx and Rx and the reflection phases of RISs, the downlink SE can be improved. We show that the optimal solution is challenging and may not satisfy the channel rank requirement. Given the potential of the flexible channel customization discussed in Section \ref{sec-3}, we propose a low-complexity solution. The required transmit power for guaranteeing $s$-stream transmission is derived as well.\par

The conventional MIMO is considered, that is, $N_{\rm R}<K<N_{\rm T}$. Supposing that the transmitted $s$-stream signal follows a Gaussian distribution, the downlink SE is then given by
\begin{equation}\label{eq-R}
\begin{aligned}
  R = {\log _2}\det ( {\bf{I}} + \frac{1}{\sigma^2}&{\bf{W}}_{{\rm{BB}}}^H{\bf{W}}_{{\rm{RF}}}^H{\bf{H}}{{\bf{F}}_{{\rm{RF}}}}{{\bf{F}}_{{\rm{BB}}}}\\
  \times&{\bf{F}}_{{\rm{BB}}}^H{\bf{F}}_{{\rm{RF}}}^H{{\bf{H}}^H}{{\bf{W}}_{{\rm{RF}}}}{{\bf{W}}_{{\rm{BB}}}} ).
\end{aligned}
\end{equation}
The SE maximization problem can be formulated as
\begin{equation}\label{eq-SS_max_ori}
  \begin{aligned}
\qquad&\mathop {\max }\limits_{{{\bf{W}}_{{\rm{RF}}}},{{\bf{W}}_{{\rm{BB}}}},\left\{ {{{\bf{\Gamma }}_k}} \right\}_{k = 1}^K,{{\bf{F}}_{{\rm{RF}}}},{{\bf{F}}_{{\rm{BB}}}}} \,R\\
{\rm{s.t. }}\quad&\qquad\left\| {{{\bf{F}}_{{\rm{RF}}}}{{\bf{F}}_{{\rm{BB}}}}} \right\|_F^2 \le E,\\
&\qquad\sqrt {{N_{\rm{T}}}} \left| {{{\bf{F}}_{{\rm{RF}}}}\left( {n,m} \right)} \right| = 1,\forall n,m,\\
& \qquad\sqrt {{N_{\rm{R}}}} \left| {{{\bf{W}}_{{\rm{RF}}}}\left( {n,m} \right)} \right| = 1,\forall n,m,\\
&\qquad{{\bf{\Gamma }}_k} = {\bf I}\otimes {\rm{diag}}\left( {{e^{j{\varpi _{k,1}}}}, \cdots ,{e^{j{\varpi _{k,{N_{{\rm{S,h,}}k}}}}}}} \right).
\end{aligned}
\end{equation}
Such an optimization problem poses challenges in obtaining the optimal solution because the objective function $R$ coupled with $K+4$ matrix variables and the constant modulus constraints on ${{\bf{F}}_{{\rm{RF}}}}$, ${{\bf{W}}_{{\rm{RF}}}}$, and ${{\bf{\Gamma }}_k}$ are non-convex. Given the reflection phases of RISs $\{ {{{\bf{\Gamma }}_k}} \}_{k = 1}^K$, the composite channel $\bf H$ is fixed, and the optimal hybrid beamforming at the Tx and Rx can be obtained via the SVD of $\bf H$ in \eqref{eq-SVD}. When $\{ {{{\bf{\Gamma }}_k}} \}_{k = 1}^K$ are given and $Q\geq s$, the optimal hybrid beamforming at the Tx and Rx can be expressed as
\begin{equation}\label{eq-opt_Tx_Rx}
\begin{aligned}
  &{{\bf{F}}_{{\rm{RF}}}}{{\bf{F}}_{{\rm{BB}}}} = {{\bf{V}}_s}{{\bf{P}}^{1/2}},\\
  &{{\bf{W}}_{{\rm{RF}}}}{{\bf{W}}_{{\rm{BB}}}} = {{\bf{U}}_s},
\end{aligned}
\end{equation}
where ${{\bf{V}}_s}$ and ${{\bf{U}}_s}$ are the first $s$ columns of $\bf V$ and $\bf U$, respectively; ${\bf{P}} = {\rm{diag}}\left( {{p_1}, \cdots ,{p_s}} \right)$ is the water-filling power allocation matrix. The diagonal element in ${\bf{P}}$ is given by
\begin{equation}\label{eq-p_i}
  {p_i} = \max{\left( {\mu  - \frac{{{\sigma ^2}}}{{{\lambda _i}}}},0 \right) },
\end{equation}
where $\mu$ is the water level that satisfies $\sum\nolimits_{i = 1}^s {{p_i}}  = E$.\par
Although the optimal hybrid beamforming at the Tx and Rx can be designed via the SVD of $\bf H$, this design faces three challenges.
\begin{itemize}
  \item Requirement for channel rank: To realize $s$-stream signal transmission, channel rank should be larger than $s$. Such value is unattainable in the conventional strong LoS mmWave communication systems. As a result of the introduction of RISs, channel rank can be customized. Thus, the challenge lies in configuring the reflection phases of RISs.
  \item Requirement for minimum transmit power: The water-filling power allocation matrix depends not only on the singular values $\{ {\sqrt{\lambda _i}} \}_{i = 1}^s$ but also on transmit power $E$. When noise power is fixed while $E$ continues to decrease, the water level $\mu$ continues dropping such that only one data stream can be allocated with power even if channel rank is larger than $s$. Therefore, the minimum transmit power should be derived to avoid zero power allocated to data streams.
  \item Decomposition of digital and analog beamforming: Although hybrid beamforming at the Tx and Rx can be easily achieved as shown in \eqref{eq-opt_Tx_Rx}, it should be  decoupled further into digital and analog beamforming in the hybrid architecture. \textcolor{black}{The manifold-based optimization method can be applied to solve this matrix decomposition problem \cite{joint_Dai}. \textcolor{black}{However, the accuracy of the reconstructed hybrid beamforming will inevitably be lost, and high computational complexity is required.}}
\end{itemize}\par
Aiming at these challenges, we attempt to satisfy the requirement for channel rank by channel customization and derive a closed-form expression for the required transmit power. On the basis of channel customization, a low-complexity design for the digital and analog beamforming is proposed without matrix decomposition.
\subsection{Joint Tx-RISs-Rx Design in the High SNR Regime}\label{sec-4.1}
In the high SNR regime, the equal power allocation is asymptotically optimal, that is, $p_i=E/s$ \cite{MIMO_OFDM}. In this case, substituting the SVD-based hybrid beamforming \eqref{eq-opt_Tx_Rx} into the SE \eqref{eq-R}, we have
\begin{equation}\label{eq-R_SVD}
  R = {\log _2}\det \left({\bf{I}} + \frac{E}{s\sigma^2}{\rm diag}({\lambda_1, \cdots,\lambda_s}) \right) .
\end{equation}
Utilizing ${\log _2}\det ({\rm diag}({x_1, \cdots,x_s}))={\sum\nolimits_{i = 1}^s {{{\log }_2}{x_i}} }$ for \eqref{eq-R_SVD}, the original optimization problem in \eqref{eq-SS_max_ori} can be expressed as
\begin{equation}\label{eq-high_SNR_SVD}
  \begin{aligned}
\mathop {\max }\limits_{\left\{ {{{\bf{\Gamma }}_k}} \right\}_{k = 1}^K}&\quad \sum\limits_{i = 1}^s {{{\log }_2}\left( {1 + \frac{E}{s\sigma^2}\lambda_i} \right)} \\
{\rm{s.t. }}\;&\quad{{\bf{\Gamma }}_k} = {\bf I}\otimes {\rm{diag}}\left( {{e^{j{\varpi _{k,1}}}}, \cdots ,{e^{j{\varpi _{k,{N_{{\rm{S,h,}}k}}}}}}} \right).
\end{aligned}
\end{equation}
In the conventional strong LoS mmWave communication system where the channel between the Tx and Rx is immutable, the objective function in \eqref{eq-high_SNR_SVD} is a constant with ${\lambda _1} \ne 0$ and ${\lambda _i} = 0$ for $i = 2, \cdots ,\min ( {{N_{\rm{T}}},{N_{\rm{R}}}} )$, which means only one data stream can be transmitted. With the help of RISs, singular values of the composite channel can be modified to establish a favorable channel for the multi-stream. However, in general cases, calculating the singular values, which is required in \eqref{eq-high_SNR_SVD}, does not have a closed-form expression with ${\{ {{{\bf{\Gamma }}_k}} \}_{k = 1}^K}$ and must be calculated numerically, e.g., through an iterative algorithm, as specified in \cite{dynamic_array} and \cite{singular_inter}. Furthermore, channel rank should be larger than $s$. Having a closed-form expression of singular values is important for the SE maximization problem. To address this challenge, we use the proposed channel customization, which provides a good approximation of singular values.\par

To provide multi-stream transmission for the Rx in a specific coverage area, we follow the channel customization scheme proposed in Section \ref{sec-3} to deploy $K$ RISs in the DFT directions of the Tx and apply the RISs segmentation algorithm to generate RISs sets ${{\mathcal K}_ \bot }$ and ${\bar{\mathcal K}_ \bot }$. Keeping $| {{\alpha _i}} | \approx 0$ for $ i \in {\bar{\mathcal K}_ \bot }$, $\{ {{\lambda _i}} \}_{i = 1}^s$ can be approximated by $\{|\alpha_k|^2, {k \in {{\mathcal K}_ \bot } \cup \{ 0 \}}\}$, and ${\bf{H}}  = {{\bf{A}}_{{\rm{R,}} \bot }}{{\bf{\Sigma }}_ \bot }{\bf{A}}_{{\rm{T,}} \bot }^H$ in \eqref{eq-H-LoS-dom} can be regarded as the SVD. On the basis of this customized composite channel, the optimal hybrid beamforming at the Tx and Rx can be easily designed without matrix decomposition, that is,
\begin{equation}\label{eq-CC_optimal_hybrid}
  \begin{aligned}
&{{\bf{F}}_{{\rm{BB}}}} = \sqrt{\frac{E}{s}}{\bf{I}},{{\bf{F}}_{{\rm{RF}}}} = {{\bf{A}}_{{\rm{T,}} \bot }},\\
&{{\bf{W}}_{{\rm{BB}}}} = {\bf{I}},{{\bf{W}}_{{\rm{RF}}}} = {{\bf{A}}_{{\rm{R,}} \bot }}.
\end{aligned}
\end{equation}
Then, \eqref{eq-high_SNR_SVD} can be rewritten as
\begin{equation}\label{eq-high_SNR_CC}
  \begin{aligned}
\mathop {\max }\limits_{\left\{ {{{\bf{\Gamma }}_k}},k \in {{\mathcal K}_ \bot } \right\}} &\sum\limits_{i \in {{\mathcal K}_ \bot } \cup \left\{ 0 \right\}} {{{\log }_2}\left( {1 + \frac{E}{{s\sigma^2}}\left| {{\alpha _i}} \right|^2} \right)}  \\
{\rm{s.t. }}\quad&\quad {{\bf{\Gamma }}_k} = {\bf I}\otimes {\rm{diag}}\left( {{e^{j{\varpi _{k,1}}}}, \cdots ,{e^{j{\varpi _{k,{N_{{\rm{S,h,}}k}}}}}}} \right).
\end{aligned}
\end{equation}
Considering that $|\alpha_k|$ is only determined by ${{\bf{\Gamma }}_k}$, \eqref{eq-high_SNR_CC} can be decoupled into $s-1$ independent sub-problems. For RIS $k\in {\mathcal K}$, the sub-problem is given by
\begin{equation}\label{eq-high_SNR_sub-CC}
  \begin{aligned}
\mathop {\max }\limits_{ {{{\bf{\Gamma }}_k}}}  &\quad{{{\log }_2}\left( {1 + \frac{E}{{s\sigma^2}}\left| {{\alpha _k}} \right|^2} \right)}  \\
{\rm{s.t. }}&\quad {{\bf{\Gamma }}_k} = {\bf I}\otimes {\rm{diag}}\left( {{e^{j{\varpi _{k,1}}}}, \cdots ,{e^{j{\varpi _{k,{N_{{\rm{S,h,}}k}}}}}}} \right).
\end{aligned}
\end{equation}
\textcolor{black}{The optimal solution for \eqref{eq-high_SNR_sub-CC} is to maximize $| {{\alpha _k}} |^2$, which can be realized by setting $| {f( {{{\bf{\Gamma }}_k}} )} |$ as the maximum according to \emph{Lemma} \ref{lemma-1}.}  \textcolor{black}{Notably, we do not claim the optimality of our joint Tx-RISs-Rx design because not all RISs are designed and the reflection matrix of RISs for channel customization may not be the optimal solution in terms of SE.\footnote{\textcolor{black}{In the proposed joint-Tx-RISs-Rx design, only $s-1$ RISs are configured for the $s$-stream transmission. When $s<{\min}\{N_{\rm T},{N_{\rm R}}\}$, more RISs can be utilized to increase channel rank further. However, the number of data streams cannot increase because $s$ is limited by $\min \{ {{K_{\rm{T}}},{K_{\rm{R}}}} \}$. \textcolor{black}{Additional RISs can improve SE in terms of SNR. However, the logarithmic scaling that SE increases logarithmically with SNR indicates a rapidly diminishing gain obtained from RISs.} To fully utilize RIS in providing multiplexing gain, the remaining $K-s+1$ RISs are supposed to be designed for other Rxs when channel customization is extended to the multi-Rx scenario.}}}. However, it exhibits the advantages of completely removing the matrix decomposition complexity of the SVD-based hybrid beamforming at the Tx and Rx. Moreover, it allows for the intuitive sensing of how RISs can change the singular values of the composite channel. The closed-form expression for singular values can also be used to determine a reasonable transmit power for the $s$-stream transmission, as shown in the succeeding subsection.\par

\subsection{Required Transmit Power for Realizing the $s$-stream Transmission}\label{sec-4.2}
In general cases, the SNR may be insufficient in guaranteeing the asymptotical optimality of equal power allocation. Following the channel customization and the SVD-based hybrid beamforming in Section \ref{sec-4.1} and replacing the equal power allocation with the optimal water-filling algorithm, \eqref{eq-high_SNR_CC} should be modified as
\begin{equation}\label{eq-low_SNR_CC}
  \begin{aligned}
\mathop {\max }\limits_{\left\{ {{{\bf{\Gamma }}_k}},k \in {{\mathcal K}_ \bot } \right\},{\bf P}}& \sum\limits_{i \in {{\mathcal K}_ \bot } \cup \left\{ 0 \right\}} {{{\log }_2}\left( {1 + \frac{\left| {{\alpha _i}} \right|^2}{\sigma^2}{p_i}} \right)}  \\
{\rm{s.t. }}\quad\;&\quad {p_i} = \max{\left( {\mu  - \frac{\sigma^2}{{{\left| {{\alpha _i}} \right|^2}}}},0 \right) },\\
&\sum\limits_{i \in {{\mathcal K}_ \bot } \cup \left\{ 0 \right\}} p_i=E,\\
&\quad {{\bf{\Gamma }}_k} = {\bf I}\otimes {\rm{diag}}\left( {{e^{j{\varpi _{k,1}}}}, \cdots ,{e^{j{\varpi _{k,{N_{{\rm{S,h,}}k}}}}}}} \right).
\end{aligned}
\end{equation}
\begin{figure}[!t]
\centering
    \includegraphics[width=0.5\textwidth]{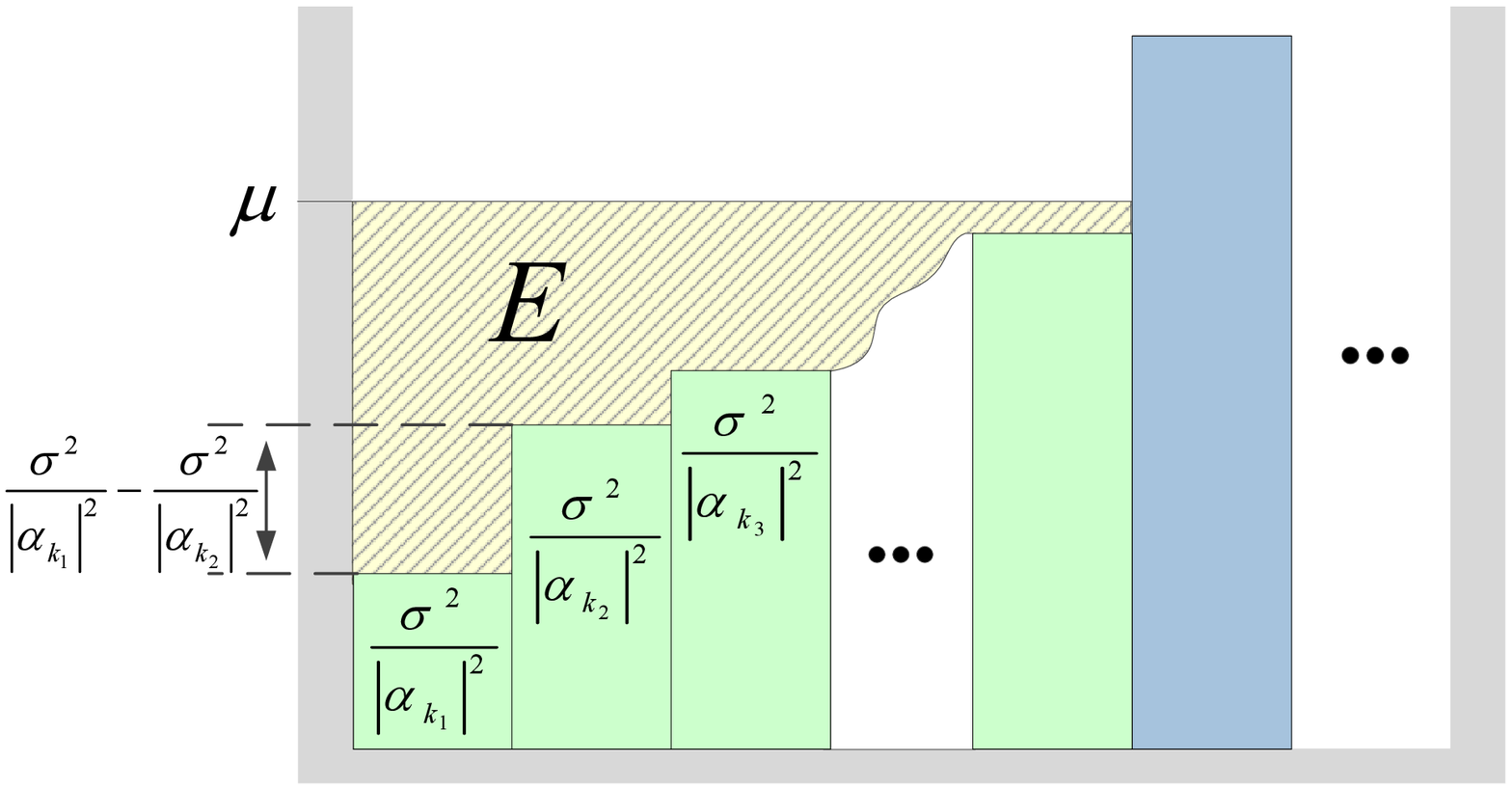}
\caption{Schematic diagram of water-filling power allocation algorithm.}
\label{Fig.water_general}
\end{figure}\par
\begin{figure}[!t]
\centering
    \includegraphics[width=0.5\textwidth]{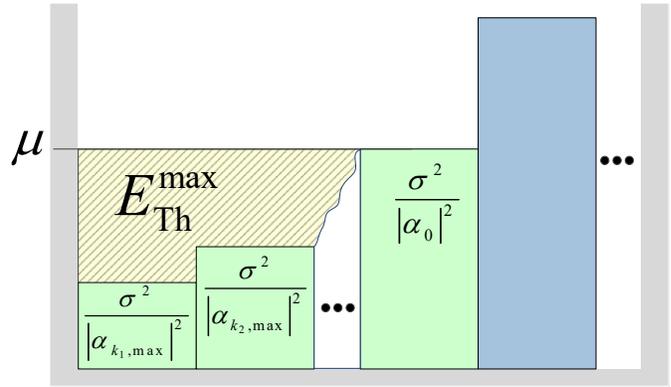}
\caption{A critical point for water-filling power allocation algorithm.}
\label{Fig.water_max}
\end{figure}

Fig. \ref{Fig.water_general} illustrates the procedure of the water-filling power allocation algorithm, which pours the ``water'' $E$ into the tank. The height from the bottom of the tank, $\sigma^2/| {{\alpha _{{k_i}}}} |^2$, to the water level $\mu$ is the power allocated to sub-channel mode $k_i$.  When $E$ continues to decrease, the bottom of the sub-channel modes will gradually break through the water lever $\mu$, thus resulting in a decreasing number of available sub-channel modes. If the first altitude intercept is larger than the total number of ``water,'' that is, $( \sigma^2/ | {{\alpha _{{k_1}}}} |^2 - \sigma^2 /| {{\alpha _{{k_2}}}} |^2 ) \ge E$, the total transmit power is allocated to only one data stream. Therefore, to realize the $s$-stream transmission, $E$ should be larger than a certain threshold ${E_{\rm Th}}$ so that the minimal power is larger than $0$, that is,
\begin{equation}\label{eq-p_min}
  {p_{\min }} = \mu  - \frac{\sigma^2}{{\left| {{\alpha _{\min }}} \right|^2}} > 0,
\end{equation}
where $| {{\alpha _{\min }}} |$ is the minimum of $| {{\alpha _i}} |$ for $i \in {{\mathcal K}_ \bot } \cup \{ 0 \}$. Considering that $N_{{\rm S},i}$ given in \eqref{eq-N_s,k_equal} enables $| {{\alpha _i}} |\geq | {{\alpha _0}} |$, $| {{\alpha _{\min }}} | = | {{\alpha _0}} |$ is assumed. In this case, according to the power constraint $\sum\nolimits_i {{p_i}} =E$, the water level can be obtained as
\begin{equation}\label{eq-mu}
  \mu  =  {\frac{{E}}{s} + \sum\limits_{i \in {{\mathcal K}_ \bot } \cup \left\{ 0 \right\}} {\frac{\sigma^2}{s{\left| {{\alpha _i}} \right|^2}}} }.
\end{equation}
\textcolor{black}{Note that the water level is implicitly affected by the quantization error of the RIS reflection phase. It can be inferred from \emph{Lemma} \ref{lemma-1} that $| {{\alpha _i}} |$ ($i \in {{\mathcal K}_ \bot }$) decreases with the quantization bits $b$, which means that the water level increases with the quantization error. Given that $| {{\alpha _0}} |$ is a constant, $p_0$ increases when the water level is elevated. Because the total power is constrained, $p_i$ ($i \in {{\mathcal K}_ \bot }$) should be decreased accordingly. In other words, the water-filling power allocation will be disturbed by the quantization error.}  Substituting \eqref{eq-mu} into \eqref{eq-p_min}, we have
\begin{equation}\label{eq-SNR_Th}
  E  > \frac{{\left( {s - 1} \right)\sigma^2}}{{\left| {{\alpha _{0 }}} \right|^2}} - \sum\limits_{i \in {{\mathcal K}_ \bot } } {\frac{\sigma^2}{{\left| {{\alpha _i}} \right|^2}}}  = {E_{{\rm{Th}}}}.
\end{equation}
{Except for $s$, the threshold $E_{\rm Th}$ is also determined by path gains $\{|\alpha_i|, {i \in {{\mathcal K}_ \bot } \cup \{ 0 \}}\}$. As a result, $E_{\rm Th}$ will vary with the reflection matrices $\{{\bf \Gamma}_i, {i \in {{\mathcal K}_ \bot } }\}$. Nevertheless, when $E$ is larger than the maximum of $E_{\rm Th}$, ${p_i}{\rm{ > 0}}$ will always hold for ${i \in {{\mathcal K}_ \bot } \cup \{ 0 \}}$ given any ${\bf \Gamma}_i$ constrained by $| {{\alpha _i}} |\geq | {{\alpha _0}} |$.  The maximized $| {{\alpha _i}} |$ turns $E_{\rm Th}$ into the maximum. Thus, the maximal $E_{\rm Th}$ can be obtained as}
\textcolor{black}{\begin{equation}\label{eq-SNR_Th_max}
  \begin{aligned}
&{E}_{{\rm{Th}}}^{\max } = \frac{\left( {s - 1} \right)\sigma^2}{{\left| {{\alpha _0}} \right|^2}} - \sum\limits_{i \in {{\mathcal K}_ \bot }} {\frac{\sigma^2}{{\left| {{\alpha _{i,\max }}} \right|^2}}} \\
 &= \frac{\sigma^2}{{I_{\rm D}^2}{ {{N_{\rm{T}}}{N_{\rm{R}}}} }}\left( {\frac{{s - 1}}{{{|g_0|^2}}} - \sum\limits_{i \in {{\mathcal K}_ \bot }} {\frac{1}{{{I_i^2}}{{N^2_{{\rm{S}},i}}|{g_{{\rm{R}},i}}{g_{{\rm{T}},i}}|^2{\rm sinc}^2 (\frac{\pi }{{{2^b}}})}}} } \right)\\
 &=  {\frac{{16\pi^2 \sigma^2}}{{I_{\rm D}^2}\lambda^2{{N_{\rm{T}}}{N_{\rm{R}}}}}} \left( {{r^2_0}\left( {s - 1} \right) - \sum\limits_{i \in {{\mathcal K}_ \bot }} {\frac{{{16{\pi^2}}{r^2_{{\rm{R}},i}}{r^2_{{\rm{T}},i}} }}{{{I_i^2}\lambda^2{N^2_{{\rm{S}},i}}{\rm sinc}^2 (\frac{\pi }{{{2^b}}})}}} } \right),
\end{aligned}
\end{equation}}
where ${{| {{\alpha _{i,\max }}} |}}$ is the maximum of $| {{\alpha _i}} |$ by setting $| {f( {{{\bf{\Gamma }}_i}} )} |$ as the maximum according to \emph{Lemma} \ref{lemma-1}. When system parameters are fixed, we can see that ${E}_{{\rm{Th}}}^{\max }$ is determined by the distances from the Rx to the Tx and to RISs. In other words, different positions in the coverage require different levels of the transmit power for the Rx to realize the $s$-stream transmission. Fig. \ref{Fig.water_max} explains why ${E}_{{\rm{Th}}}^{\max }$ enables ${p_i}{\rm{ > 0}}$ for ${i \in {{\mathcal K}_ \bot } \cup \{ 0 \}}$ given any ${\bf \Gamma}_i$ constrained by $| {{\alpha _i}} |\geq | {{\alpha _0}} |$. When $E={E}_{{\rm{Th}}}^{\max }$ and ${\bf \Gamma}_i$, $i\in {\mathcal K}_\bot$, is configured to achieve $| {{\alpha _i}} |={{| {{\alpha _{i,\max }}} |}}$, the water level will reach exactly the bottom of the sub-channel mode that corresponds to the direct link between the Tx and Rx. In this case, arbitrary modifications on the ${\bf \Gamma}_i$ will lead to $|{{\alpha _i}}|<{{| {{\alpha _{i,\max }}} |}}$, thus raising the bottom of its corresponding sub-channel mode. Once the bottom of the effective sub-channel mode rises, the water level must be elevated to overwhelm the bottom ${\sigma^2}/{{| {{\alpha _0}} |^2}}$, which means ${p_i}{\rm{ > 0}}$ holds for ${i \in {{\mathcal K}_ \bot } \cup \{ 0 \}}$.

\par

In the following, $c\in {\bar{\mathcal K}_\bot}$ is assumed to not break the orthogonality of ${\bf A}_{{\rm R},\bot}$ when it is enrolled into ${{\mathcal K}_\bot}$. Then, when another data stream is demanded, we can set $| {{\alpha _c}} |={{| {{\alpha _{c,\max }}} |}}$ to generate an additional sub-channel mode with the bottom being
${\sigma^2}/{{| {{\alpha _{c,\max}}} |^2}}$. In this case, the difference between the maximal thresholds for $s+1$ and $s$ data streams satisfies
\textcolor{black}{\begin{equation}\label{eq-SNR_Th_max_diff}
\begin{aligned}
\Delta E_{{\rm{Th}}}^{\max }&=E_{{\rm{Th}}}^{\max }{|_{s + 1}} - E_{{\rm{Th}}}^{\max }{|_s}\\
&=  {\frac{{16\pi^2 \sigma^2}}{{I_{\rm D}^2}\lambda^2{{N_{\rm{T}}}{N_{\rm{R}}}}}} \left( {{r^2_0} - \frac{{{16{\pi^2}}{r^2_{{\rm{R}},c}}{r^2_{{\rm{T}},c}} }}{{I_c^2}\lambda^2{N^2_{{\rm{S}},c}}{\rm sinc}^2 (\frac{\pi }{{{2^b}}})}} \right).
\end{aligned}
\end{equation}}
This extra transmit power required for another data stream depends on the chosen RIS $c$ and the position of the Rx because they determine $N_{{\rm S},{c}}$, $r_{{\rm R,}c}$, and $r_0$. When the scale of RIS $c$, $N_{{\rm S},{c}}$, is increased, the demand for  extra transmit power increased. In the extreme case where $N_{{\rm S},{c}}\to \infty$, we have $\Delta E_{{\rm{Th}}}^{\max }\to \sigma^2/|\alpha_0|^2$. Fig. \ref{Fig.water_max} can explain this phenomenon as well. Digging up another sub-channel mode that has a bottom ${\sigma^2}/{{| {{\alpha _{c,\max}}} |^2}}$ below the original water level will inevitably bring down the water level and expose the bottom of the highest sub-channel mode to the air. To overwhelm the exposed sub-channel mode again, the water should increase accordingly. Considering that increasing $N_{{\rm S},{c}}$ will depress the bottom ${\sigma^2}/{{| {{\alpha _{c,\max}}} |^2}}$, this water increment will increase with $N_{{\rm S},{c}}$ and reach its limited imposed by $\sigma^2/|\alpha_0|^2$.\par

In the above scenario, another data stream is obtained at the cost of the transmit power increment in the condition that $| {{\alpha _i}} |={{| {{\alpha _{i,\max }}} |}}$ for $i\in {\mathcal K}_\bot$. This result is common in a traditional water-filling algorithm. However, with the channel customization ability of RISs, we can transmit additional data streams without enlarging the transmit power for the water-filling algorithm. Different from the conventional system where the bottom of the sub-channel is uncontrollable, RISs can individually elevate the bottoms so that when another sub-channel is dug out, the water level can be pushed back to the original position without pouring additional water into the tank. From this point of view, customizing a composite channel with homogenous singular values, or a minimized truncated condition number, is the ideal condition for guaranteeing the multi-stream transmission. However, this strategy fails to achieve the maximum singular value for some sub-channels, thus inevitably resulting in a deteriorative SE, as will be illustrated in Section \ref{sec-5}.

\section{Numerical Results}\label{sec-5}
\begin{figure}[!t]
\centering
    \includegraphics[width=0.5\textwidth]{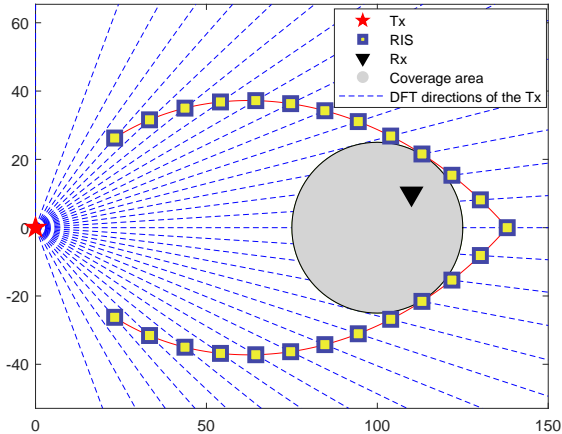}
\caption{System deployment (top view).}
\label{Fig.deployment}
\end{figure}\par

In this section, numerical results are presented to demonstrate the effectiveness of the proposed channel customization and joint Tx-RISs-Rx design in the rank deficient strong LoS mmWave communication system. The conventional MIMO is considered, that is, $N_{\rm R}<K<N_{\rm T}$. The system carrier frequency is $f_{\rm c}=28$ GHz. As shown in Fig. \ref{Fig.deployment}, the Tx, equipped with ${N_{\rm{T}}} = 32$ antennas and ${K_{\rm{T}}} = 8$ RF chains, is located in the original Cartesian coordinates ${{\bf{e}}_{\rm{T}}} = {[ {0,0} ]^T}$. \textcolor{black}{The distance in Fig. \ref{Fig.deployment} is presented in meters.} DFT directions of the Tx are represented by a series of rays emitted from ${{\bf{e}}_{\rm{T}}}$ with the angle being $\xi \in \{ \arcsin ({\frac {{2 i}}{{{N_{\rm{T}}}}} - 1 }), i=1,\cdots,N_{\rm T} \}$. \textcolor{black}{The Rx is assumed to be randomly distributed in a circular coverage with the center and the radius being ${{\bf{e}}_{\rm{C}}} = {[ {100,0} ]^T}$ and ${r_{\rm{C}}} = 25$, respectively.} The Rx is equipped with $N_{\rm R}=8$ antennas and $K_{\rm R}=s$ RF chains. Twenty-five RISs with reflection phase quantized by $b=4$ bits are mounted in the system. To deploy these RISs, a curve $\mathcal S$ that has 25 intersections with the DFT directions is provided in Fig. \ref{Fig.deployment} as an example. \textcolor{black}{The coordinates of $\mathcal S$ are given by ${\bf e}_{\mathcal S}=[\rho\cos\varepsilon,\rho\sin\varepsilon]^T$ for $\varepsilon  \in [ {-{\varepsilon _{\max }},{\varepsilon _{\max }}} ]$, where ${\varepsilon _{\max }} = \arcsin ( (K+1)/N_{\rm T} )$ is the angle of the first intersection and ${\rho } = 35 + {{( {138 - 35} )}}(1-|\varepsilon|/{{{\varepsilon _{\max }}}})$ is the distance from ${\bf e}_{\mathcal S}$ to the Tx.} According to Section \ref{sec-3.3}, we place $25$ RISs at these intersections, respectively. \textcolor{black}{Given the fixed positions of the Tx and RISs, along with the range of coverage, the number of elements required for RISs to combat  path loss is given by \eqref{eq-N_s,k_equal} with $I_k$ = 0.01 for $\forall k$. In particular, when counting clockwise from the rightmost RIS in Fig. \ref{Fig.deployment}, the element numbers of RISs are 1360, 1147, 993, 904, 875, 880, 893, 900, 890, 858, 800, 710, 582, 1147, 993, 904, 875, 880, 893, 900, 890, 858, 800, 710, and 582. } The AoA and AoD of the LoS paths are determined by the locations of the Tx, Rx, and RISs {while those of the NLoS paths are randomly distributed in ${\mathcal U}[0,\pi]$}. \textcolor{black}{The Tx and Rx are equipped with a half-wavelength antenna array, that is, $d_{\rm T}=d_{\rm R}=\lambda/2$, where $\lambda$ is the wavelength. Considering that the RIS is composed of metamaterials with sub-wavelength spacing \cite{sub-wave-1}, we set $d_{\rm S}=\lambda/8$.} \textcolor{black}{The noise power is set as $\sigma^2= -100$ dBm.} \par

Considering that the channel customization is purpose-oriented, in the following, we focus on two scenarios where the reflection phases of RISs in ${\mathcal K}_\bot$ ($|{\mathcal K}_\bot|=s-1$) are designed to achieve homogeneous path gains (HPG) to minimize the truncated condition number $T_s$ or maximal path gains (MPG) to maximize the SE $R$. More specifically, HPG and MPG indicate $\{|\alpha_i |=|\alpha_0|,|\alpha_j |\approx 0\}$ and $\{|\alpha_i |=|\alpha_{i,{\max}}|,|\alpha_j |\approx 0\}$, respectively, for $i\in {\mathcal K}_\bot$, $j\in {\mathcal{K}}\backslash{\mathcal K}_\bot$. The undesigned reflection phases that result in random path gains (RPG) are considered the worst case when the SE is assessed.
\subsection{Channel Customization}\label{sec-5.1}
\begin{figure}[!t]
\centering
    \includegraphics[width=0.5\textwidth]{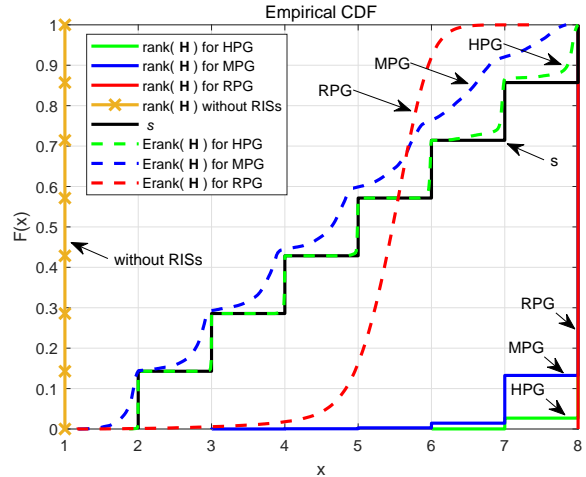}
\caption{CDF of the rank and effective rank that has been customized according to $s$.}
\label{Fig.Erank_CDF}
\end{figure}
In this subsection, we evaluate the channel customization in terms of the effective rank and the truncated condition number. Assuming that the number of data streams $s$ increases from $2$ to $8$, we plot in Fig. \ref{Fig.Erank_CDF} the CDF of the rank and effective rank of the composite channel that has been customized according to $s$. For each $s$, the Monte Carlo simulation is performed for 10,000 channel realizations by randomly placing the Rx in the coverage. The solid lines except for the stair curve for $s$ are the CDF of channel rank, which is the number of non-zero singular values, and calculated by MATLAB. Meanwhile, the dashed lines are the CDF of the effective rank determined by \eqref{eq-Erank}.\par
\begin{figure}[!t]
\centering
    \includegraphics[width=0.5\textwidth]{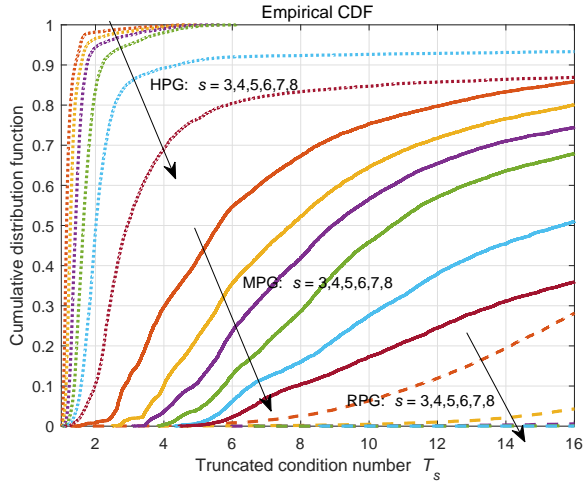}
\caption{CDF of $T_s$ for the composite channel that has been customized according to $s$.}
\label{Fig.TCN}
\end{figure}

From Fig. \ref{Fig.Erank_CDF}, we see that the rank for RPG has always been $8$, thus indicating a full column rank for the composite channel. However, this full column rank is deceptive because the effective rank for RPG is mainly distributed between $5$ and $6$. In $70,000$ channel realizations, when the reflection phases are undesigned, over $80\%$ of channels have an effective rank that is larger than $5$, thus showing an excellent channel improvement at first sight because channel rank is $1$ in the conventional strong LoS mmWave system without RISs. Although the RPG offers a high effective rank, Fig. \ref{Fig.TCN} demonstrates that its truncated condition number is large as well. Consequently, RPG will not produce a satisfactory SE, as will be demonstrated in Fig. \ref{Fig.diff_power}. Moreover, the effective rank cannot be customized according to the number of data streams once RISs are undesigned. When path gains are designed to be homogeneous, the CDF of the effective rank for HPG is matched well with that of $s$ in the low $s$ regime. In other words, the proposed channel customization scheme can shape the composite channel to have an effective rank that is almost identical to the number of data streams. When $s$ increases to $7$ and $8$, additional RISs are involved to provide extra sub-channel modes. In this case, the orthogonality of each sub-channel mode will be gradually impaired, thus resulting in a slightly inferior effective rank to the corresponding $s$. Considering that the effective rank is maximized by equal singular values, the CDF of the effective rank for MPG is more distorted when the path gains are optimized to be maximal but not homogeneous.\par

Fig. \ref{Fig.TCN} shows the CDF of the truncated condition number, $T_s$, for the customized composite channel when $s$ increases from $3$ to $8$. When the reflection phases of RISs are undesigned, $T_s$ in most channel realizations is large. For example, the presented CDF for RPG demonstrates that more than $70\%$ of channel realizations have $T_s>16$ even when $s$ is reduced to $3$. Therefore, we can conclude that the channel for RPG is poorly conditioned for the purpose of communication. When the composite channel is customized according to $s$ and the path gain of RIS $k$, $k\in {\mathcal K}_\bot$, is configured as maximum, the CDF for MPG improves dramatically. Specifically, the probability for $T_3<16$ is promoted to over $85\%$. Although CDFs for MPG drop when $s$ increases, they are still superior to that for RPG. When the path gains of the \textcolor{black}{\emph{activated}} RISs in ${\mathcal K}_\bot$ are designed to be homogeneous, $T_s$ can be reduced well, as can be seen from CDFs for HPG. When $s$ equals the maximal rank, that is $8$, the probability for $T_8>16$ is cut down to less than $15\%$. Notably, when $s\le 6$, the probability for $T_s\le 8$ is nearly $100\%$, which means that the composite channel is conditioned well.

\begin{figure}[!t]
	\centering
	\subfigure[$T_2$ for each location in the coverage]{
		\label{Fig.RPG_2_all} 
		\includegraphics[width=0.233\textwidth]{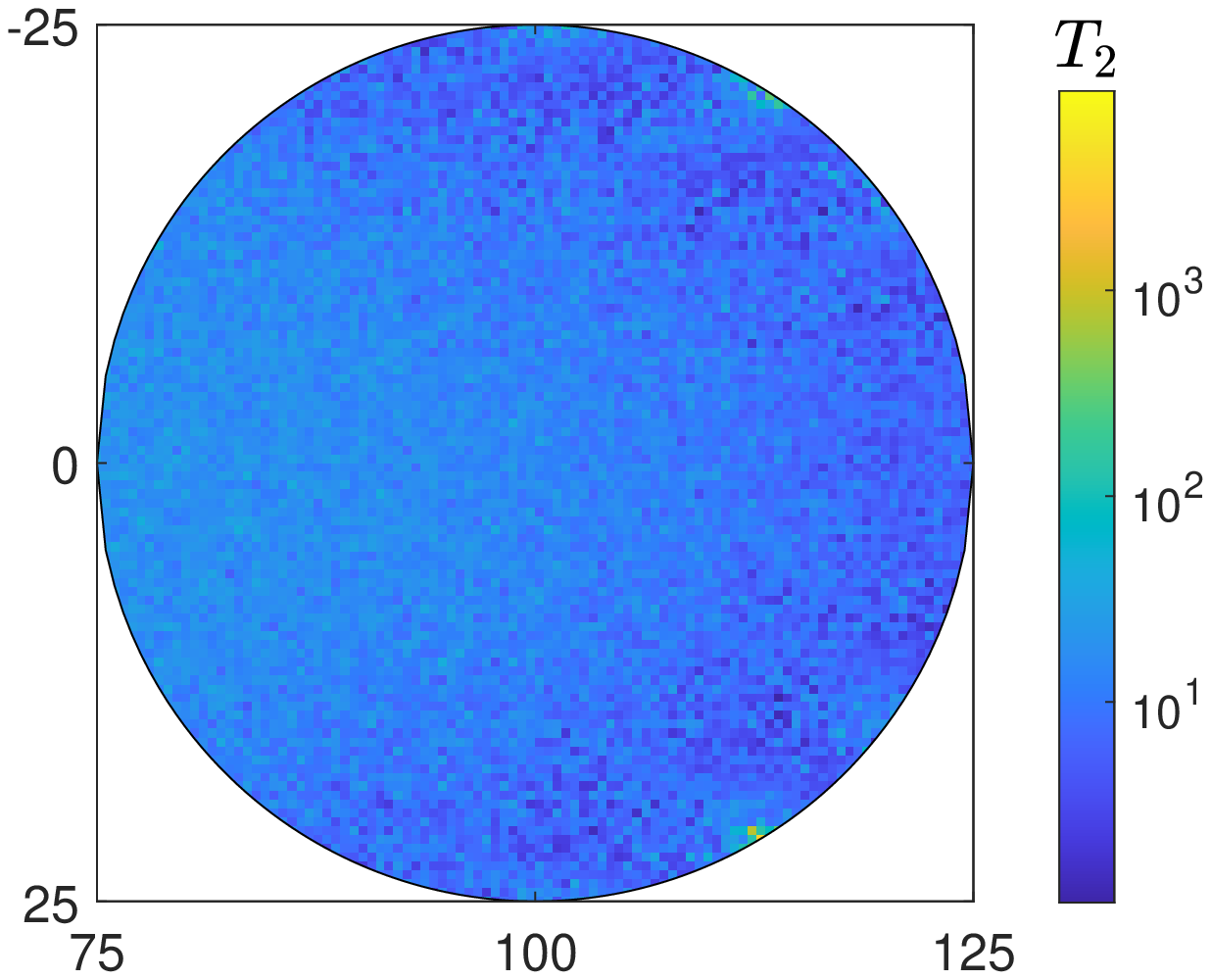}}
	\subfigure[$T_4$ for each location in the coverage]{
		\label{Fig.RPG_4_all} 
		\includegraphics[width=0.233\textwidth]{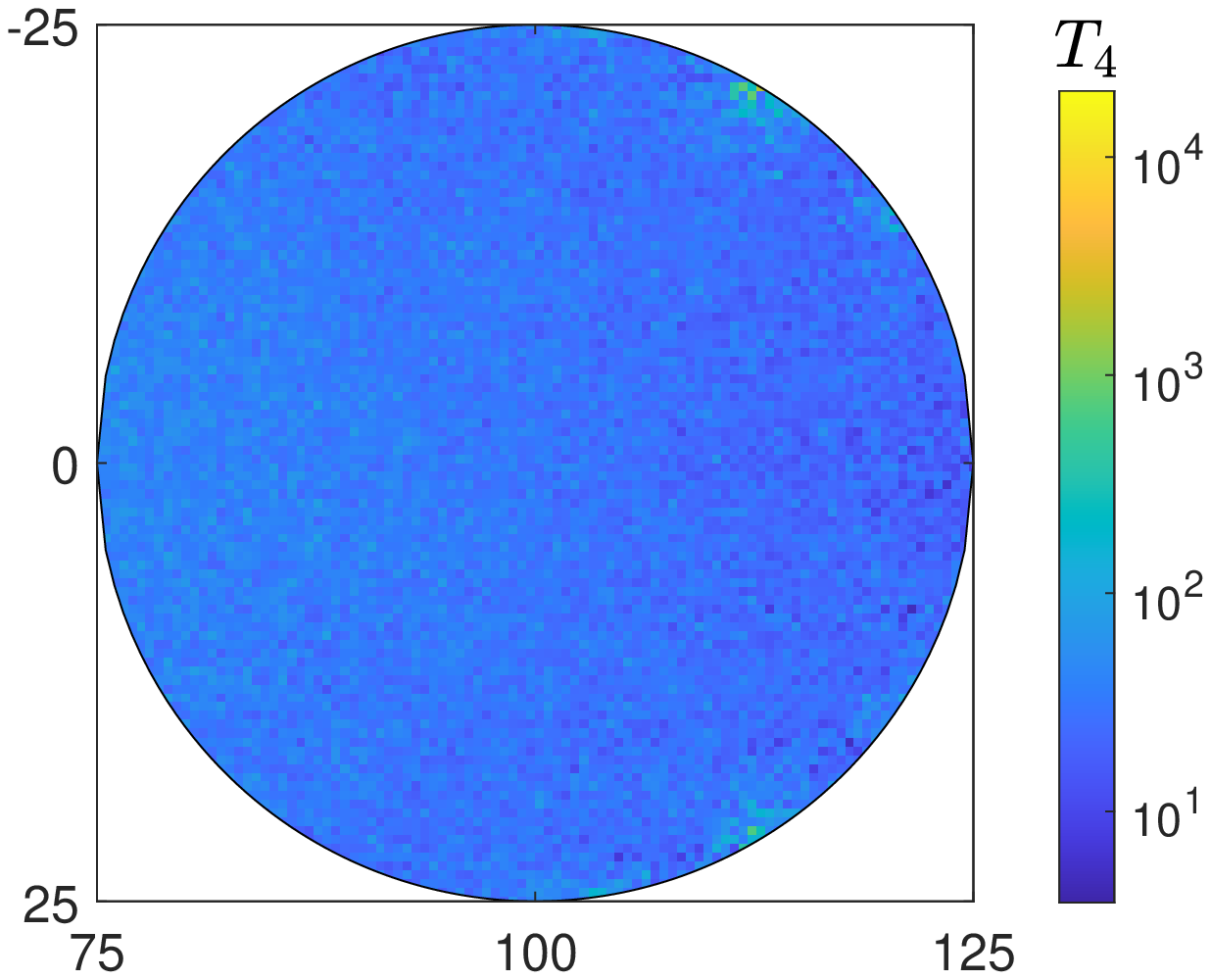}}
	\subfigure[Counterpart of (a) by removing $T_2>8$]{
		\label{Fig.RPG_2_par} 
		\includegraphics[width=0.233\textwidth]{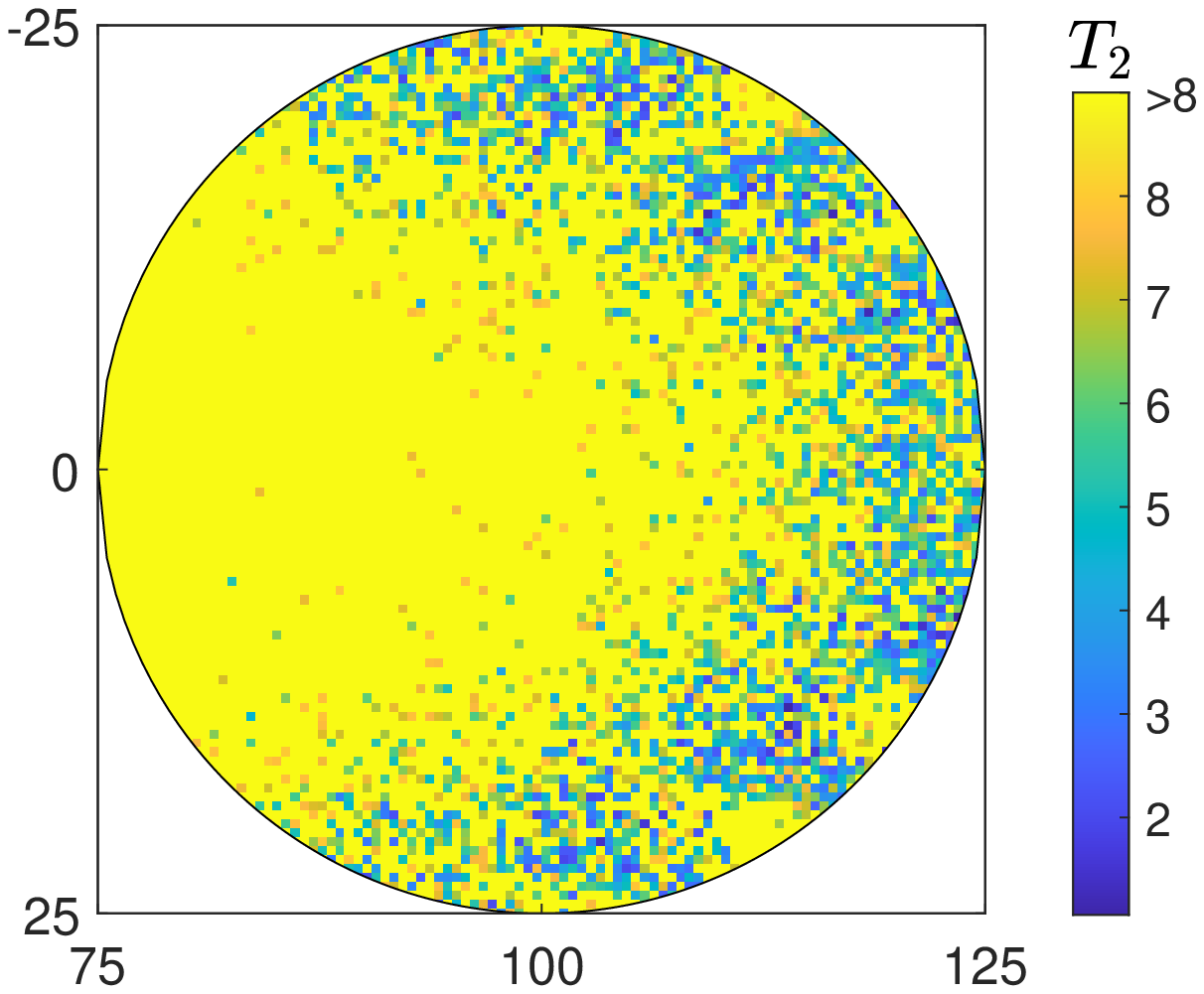}}
	\subfigure[Counterpart of (b) by removing $T_4>8$]{
		\label{Fig.RPG_4_par} 
		\includegraphics[width=0.233\textwidth]{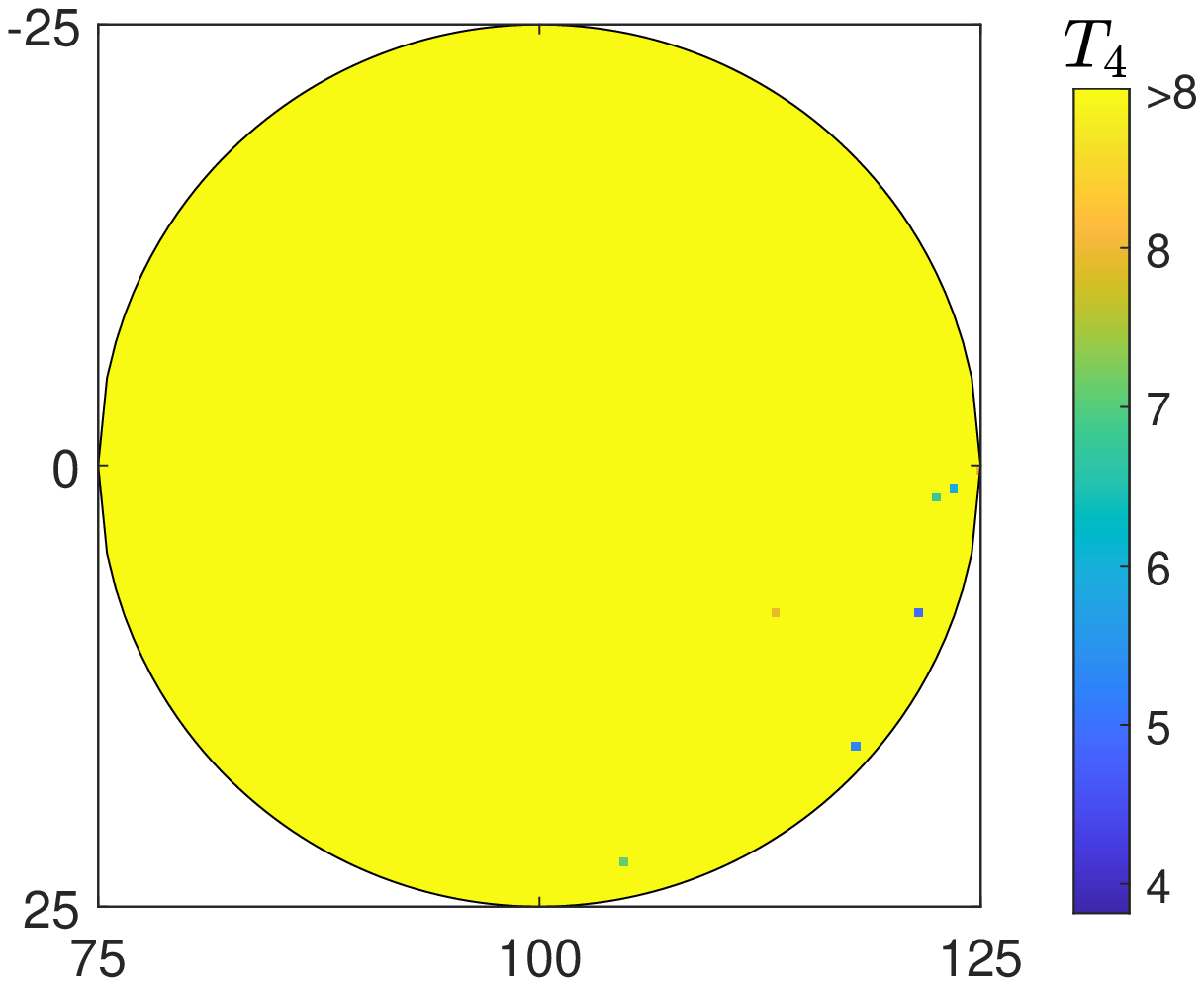}}
	\caption{Distribution of $T_s$ in the coverage for RPG.}
	\label{Fig.RPG} 
\end{figure}

Fig. \ref{Fig.RPG} illustrates the distribution of the truncated condition number in the circular coverage when the reflection phases of RISs are undesigned. The left and right sub-figures are presented with $s=2$ and $s=4$, respectively. The upper sub-figures show the full range of $T_s$, while the bottom sub-figures only present $T_s\le 8$, which is defined as well-conditioned in the following. Fig. \ref{Fig.RPG_2_all} and \ref{Fig.RPG_4_all} demonstrates that $T_s$ distributes uniformly in the coverage and the distribution is similar for different $s$. Given that most $T_2$ and $T_4$ range from $10$ to $100$, we know that only one large singular value exists and the rest of singular values are small and in the same order of magnitude. To display the coverage with the well-conditioned channel, Fig. \ref{Fig.RPG_2_par} and \ref{Fig.RPG_4_par} remove $T_s>8$. When $s=2$, a few points remain with the well-conditioned channel. However, these points rapidly disappear when $s$ increases to $4$, thus indicating that the well-conditioned channel with a high rank cannot be customized with the undesigned RISs.\par

\begin{figure}[!t]
\centering
\subfigure[$T_2$ for each location in the coverage]{
\label{Fig.HPG_2_all} 
\includegraphics[width=0.23\textwidth]{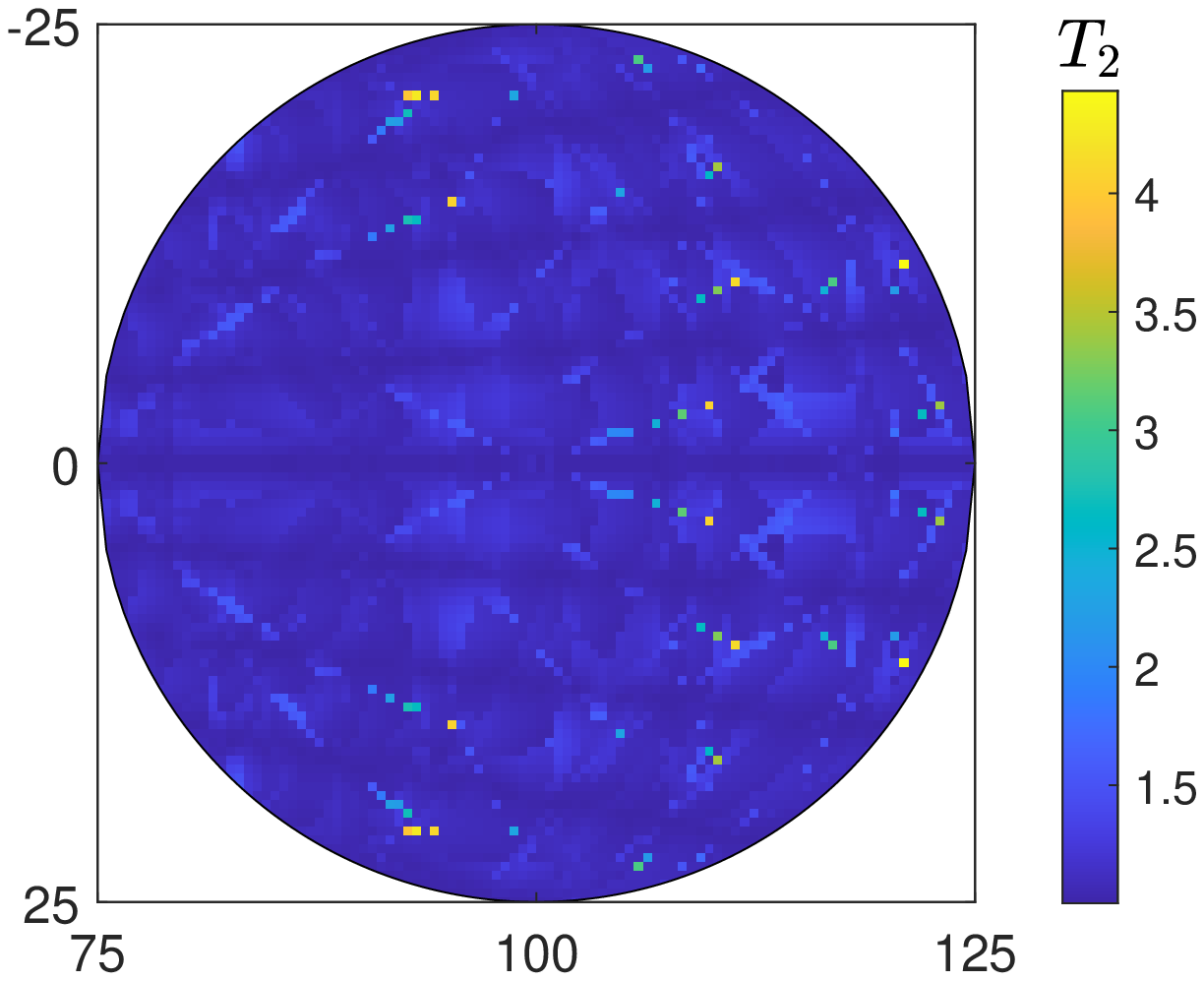}}
\subfigure[$T_4$ for each location in the coverage]{
\label{Fig.HPG_4_all} 
\includegraphics[width=0.23\textwidth]{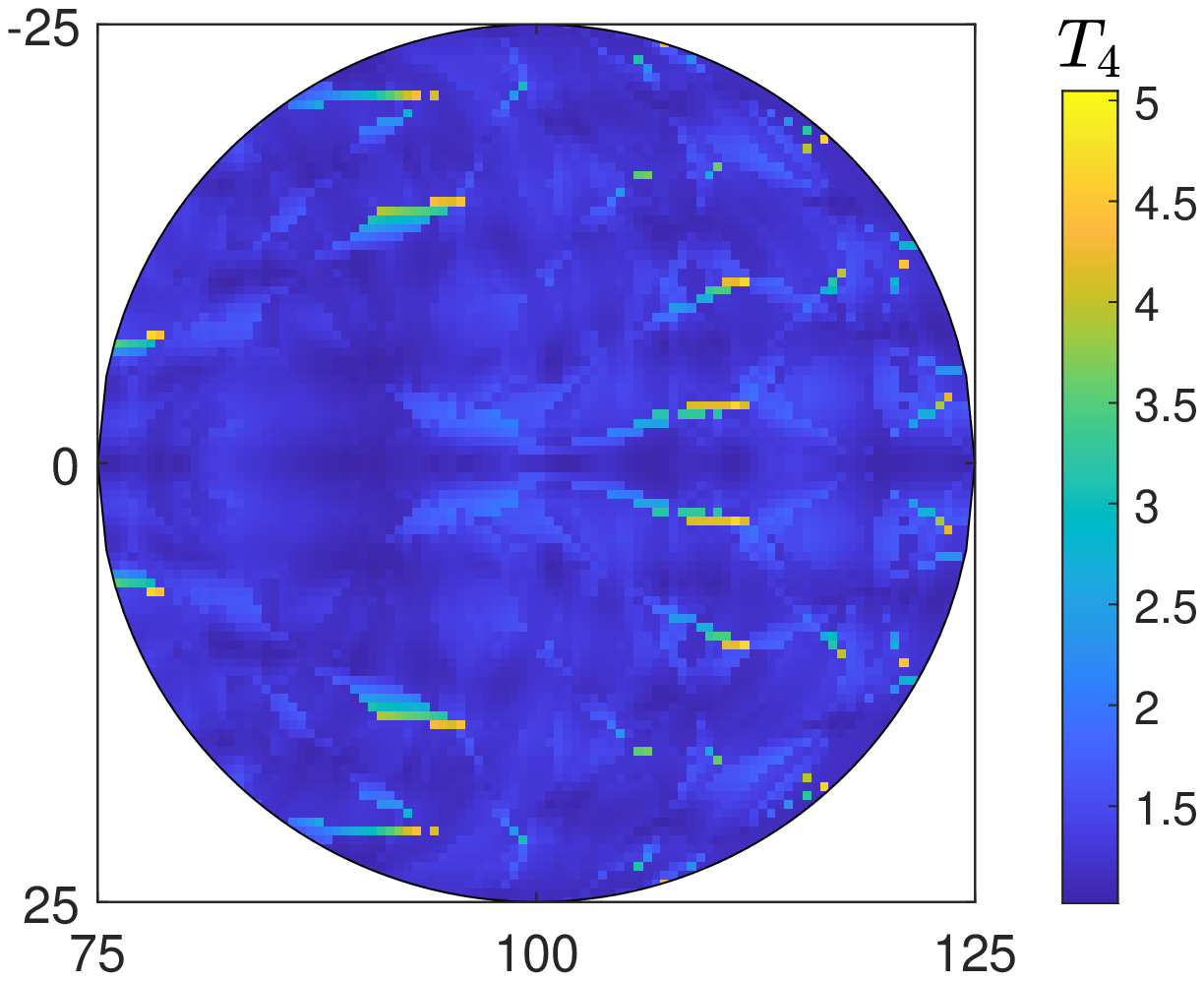}}
\caption{Distribution of $T_s$ in the coverage for HPG.}
\label{Fig.HPG} 
\end{figure}

In Fig. \ref{Fig.HPG}, channel rank is customized according to the required number of data streams, and the path gains are tuned to be homogeneous. Compared with Fig. \ref{Fig.RPG}, the range of $T_s$ decreases dramatically over the entire coverage. In addition, the well-conditioned channel can be established in every corner of the coverage. Thus, the proposed channel customization scheme is validated. More importantly, this desired channel is robust against the increasing $s$. For example, the maximal $T_s$ only increases from $4.5$ to $5$, as can be seen in Fig. \ref{Fig.HPG_2_all} and \ref{Fig.HPG_4_all}.
\subsection{Joint Tx-RISs-Rx Design}\label{sec-5.2}

\begin{figure}[!t]
\centering
\subfigure[]{
\label{Fig.equal_power} 
\includegraphics[width=0.5\textwidth]{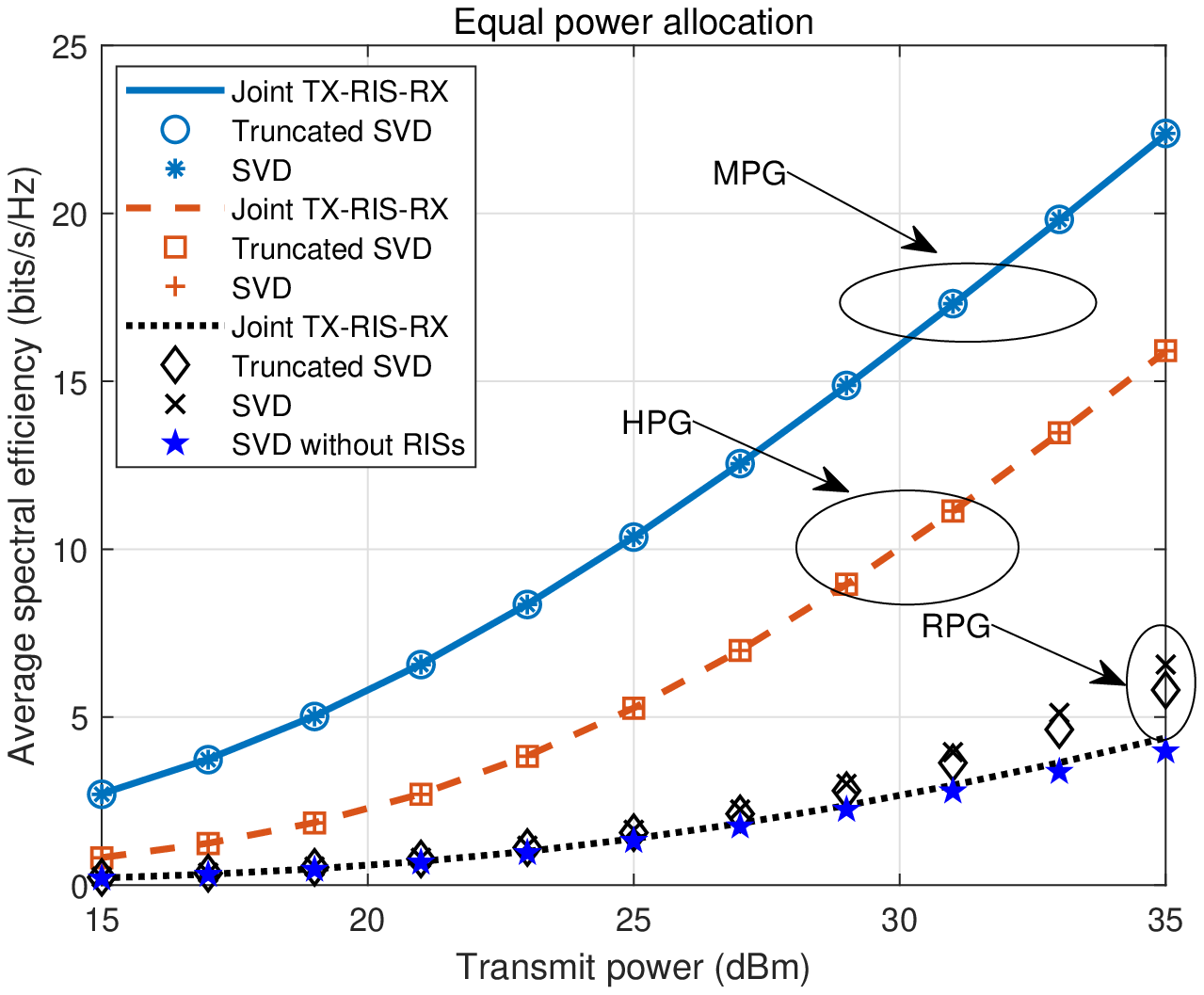}}
\subfigure[]{
\label{Fig.water_power} 
\includegraphics[width=0.5\textwidth]{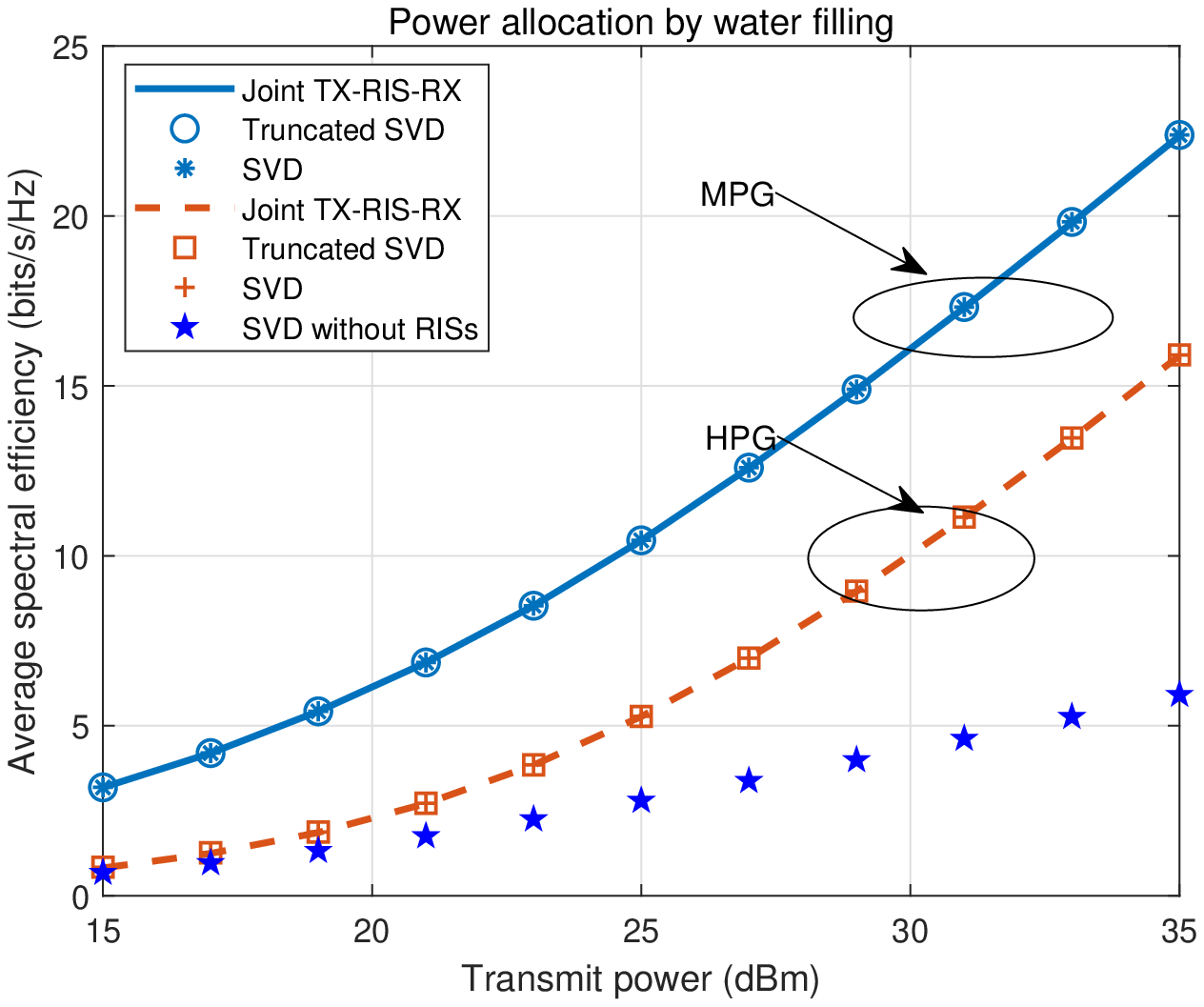}}
\caption{Average SE versus transmit power for different RISs design with (a) equal power allocation and (b) water-filling power allocation.}
\label{Fig.diff_power} 
\end{figure}
In this subsection, the proposed joint Tx-RISs-Rx design is investigated in terms of the SE averaged over the entire coverage area. The SVD- and truncated SVD-based beamforming at the Tx and Rx are considered comparisons. In the following figures, the legends \textbf{SVD} and \textbf{Truncated SVD} indicate that beamforming matrices at the Tx and Rx are designed as $\{{{\bf{F}}_{{\rm{RF}}}}{{\bf{F}}_{{\rm{BB}}}} = {{\bf{V}}}{{\bf{P}}^{1/2}},{{\bf{W}}_{{\rm{RF}}}}{{\bf{W}}_{{\rm{BB}}}} = {{\bf{U}}}\}$ and \eqref{eq-opt_Tx_Rx}, respectively. The difference is that the \textbf{SVD} does not consider the limit on the number of RF chains as the \textbf{Truncated SVD} does. Notably, the matrix decomposition for digital and analog beamforming is not considered for the \textbf{SVD} and \textbf{Truncated SVD}.\par

Fig. \ref{Fig.diff_power} demonstrates the average SE versus transmit power for different RIS designs when $s=4$. The RPG cases are not considered in Fig. \ref{Fig.water_power} because the singular values needed for water-filling algorithm can not be expressed in closed-form of RISs' reflection matrix. The high transmit power is set to combat the path loss. Evidently, the introduction of RISs brings substantial gains, especially in MPG cases, over the conventional systems without RISs. In Fig. \ref{Fig.equal_power}, when the reflection phases of RISs are undesigned, the SVD-based scheme achieves the best performance, whereas our proposal is the worst. In RPG, the composite channel is not rectified well for multi-stream transmission. Hence, even when hybrid beamformings at the Tx and Rx focus on the orthogonal RISs, the transmitted energy is still dispersed. Although the effective rank for RPG is impressive in Fig. \ref{Fig.Erank_CDF}, the average SE of RPG is the worst compared with that of MPG and HPG, thus validating our conclusion that the high effective rank offered by the undesigned RISs cannot produce a satisfactory communication performance. When the composite channel is carefully customized according to the data stream requirement, the performances of the joint Tx-RISs-Rx can be remarkably improved, regardless of power allocation schemes. Moreover, our proposal achieves almost identical SE to the SVD-based scheme without further matrix decomposition for digital and analog beamforming. Although HPG can establish a well-conditioned channel as shown in Fig. \ref{Fig.HPG}, the SE is inferior to that of MPG. \textcolor{black}{This difference is a consequence of the reduction in the array gains of \textcolor{black}{activated} RISs when their corresponding path gains are forced to be homogeneous. Given that HPG activates fewer RIS elements, it consumes less energy and may be preferred when energy efficiency performance is considered.}

\begin{figure}[!t]
\centering
    \includegraphics[width=0.5\textwidth]{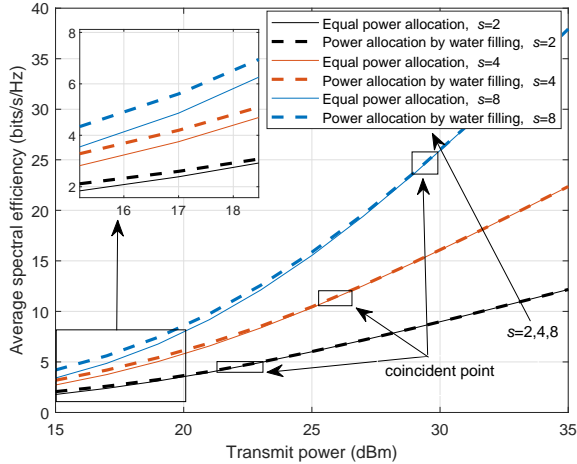}
\caption{Comparisons of the average SE for equal power allocation and water-filling power allocation with increasing $s$ when MPG is considered.}
\label{Fig.equal_water}
\end{figure}

Fig. \ref{Fig.equal_water} compares the average SE of equal power allocation and water-filling power allocation with increasing $s$ when MPG is considered. Evidently, the increment of $s$ can promote the average SE because additional sub-channel modes are excavated for multi-stream transmission. \textcolor{black}{SE improvement with more RISs indicates the SE loss incurred by the selected $s-1$ RISs designed for the proposed channel customization compared with the optimal solution that configures all RISs.} Considering that the water level must not decrease to ensure the equal power allocation is asymptotically optimal, the transmit power should increase to compensate for the water decline when another sub-channel mode that has a lower bottom than the direct link is introduced. As a result, the coincident point for equal power allocation and water-filling power allocation shifts to higher transmit power.

\subsection{Effect of NLoS Paths}\label{sec-5.3}

\begin{figure}[!t]
\centering
    \includegraphics[width=0.5\textwidth]{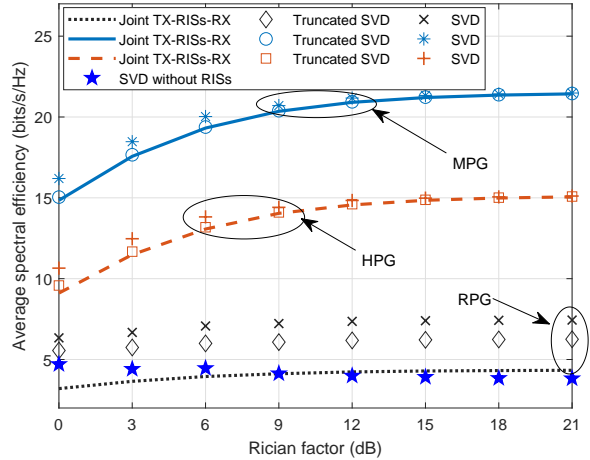}
\caption{Average SE versus Rician factor for different RISs design.}
\label{Fig.Rician}
\end{figure}

In this subsection, the effect of NLoS paths on the proposed joint Tx-RISs-Rx design is investigated by relaxing the infinite Rician factor assumption. Considering the sparse nature of mmWave channels, the number of NLoS paths is set to $3$ for ${{\bf{H}}_{{\rm{T}},k}}$, ${{\bf{H}}_{{\rm{R}},k}}$, and ${{\bf{H}}_{\rm{D}}}$. We can safely assume that the power of the LoS component is larger than that of NLoS components in the strong LoS mmWave systems. Hence, in Fig. \ref{Fig.Rician} we start the Rician factor from $0$ dB. When the composite channel is customized well by MPG or HPG, the average SE of our proposal is slightly lower than the optimal SVD-based scheme at first but the performance gap is rapidly narrowed with the increasing Rician factor. Given that the limit on RF chains is not considered in the SVD-based scheme, our proposal still has a satisfactory performance because it matches well with the truncated SVD-based scheme. This well-matched performance elucidates that the MPG and HPG are robust against NLoS paths in mmWave systems. Notably, when RISs are undesigned, our proposal is invalid in the scenario where the powers of the LoS component and NLoS components are comparative. This phenomenon emphasizes the importance of the RISs design in customizing the composite channel with strong NLoS paths. Nevertheless, the SVD- and truncated SVD-based scheme in the RISs-assisted system retain considerable performance gains than the counterpart without RISs, regardless of how RISs are designed.


\section{Conclusion}\label{sec-6}

In this study, we investigated the channel customization and joint Tx-RISs-Rx design in the hybrid mmWave communication system assisted by multiple RISs. Sparse propagation paths in the conventional mmWave system result in a rank-deficient channel that is unpopular for multi-stream transmission. To overcome this intrinsic limitation, RISs were introduced to the customization of a favorable composite channel by reconfiguring its characteristics. Different from existing works, we provided a closed-form expression that bridges the singular value of the composite channel and the reflection phase of RISs to instruct the customization process. Furthermore, a low-complexity joint Tx-RISs-Rx design was proposed, which guaranteed the required effective channel rank for multi-stream transmission by configuring the reflection phases of RISs and performing the SVD-based hybrid beamforming without matrix decomposition. Numerical results demonstrated that the effective channel rank can be accurately and flexibly customized according to the number of data streams. In terms of the average spectral efficiency, the joint Tx-RISs-Rx design was shown to be able to obtain significant gains over the conventional system without RISs.

\appendices
\section{}\label{App:A}
When the Tx, RIS, and Rx are at the same altitude, that is, ${\Phi _{{\rm{R}},k}^{\rm{D}}}={\Phi _{{\rm{T}},k}^{\rm{A}}}$, according to the definition of array gain of RIS $k$, we have
\begin{equation}\label{eq-f_defi}
	\begin{aligned}
		f\left( {{{\bf{\Gamma }}_k}} \right) &\buildrel \Delta \over = {{N_{{\rm{S}},k}}{\bf{a}}_{{\rm{S}},k}^H( {\Phi _{{\rm{R}},k}^{\rm{D}}},{\Theta _{{\rm{R}},k}^{\rm{D}}} ){{\bf{\Gamma }}_k}{{\bf{a}}_{{\rm{S}},k}}({\Phi _{{\rm{T}},k}^{\rm{A}}}, {\Theta _{{\rm{T}},k}^{\rm{A}}} )}\\
		&={{N_{{\rm{S,v,}}k}}}\sum\limits_{n = 1}^{{N_{{\rm{S,h,}}k}}} {{e^{j\left( {{\varpi _{k,n}} - \left( {n - 1} \right)\left( {\Theta _{{\rm{R}},k}^{\rm{D}} - \Theta _{{\rm{T}},k}^{\rm{A}}} \right)} \right)}}}.
	\end{aligned}
\end{equation}
It is easy to find that when the reflection phase of RISs can be continuously controlled,
\begin{equation}\label{eq-optimal_phase_c}
	\varpi _{k,n}^{{\rm{c,opt}}} = \left( {n - 1} \right)\left( {\Theta _{{\rm{R}},k}^{\rm{D}} - \Theta _{{\rm{T}},k}^{\rm{A}}} \right)
\end{equation}
is the optimal reflection phase that turns $ {f( {{{\bf{\Gamma }}_k}} )} $ to be the maximum, that is, $N_{{\rm S},k}$. However, when the resolution of the reflection phase is constrained by $b$ bits, the optimal discrete solution $\varpi _{k,n}^{{\rm{d,opt}}}$ satisfies
\begin{equation}\label{eq-optimal_phase_d}
	- \frac{\pi }{{{2^b}}} \le \varpi _{k,n}^{{\rm{d,opt}}} - \varpi _{k,n}^{{\rm{c,opt}}} \le \frac{\pi }{{{2^b}}}.
\end{equation}
When ${N_{{\rm{S,h}},k}} \to \infty $, $\{ {\varpi _{k,n}^{{\rm{d,opt}}} - \varpi _{k,n}^{{\rm{c,opt}}}} \}_{n = 1}^{{N_{{\rm{S,h}},k}}}$ will be uniformly distributed in $[ { - \frac{\pi }{{{2^b}}},\frac{\pi }{{{2^b}}}} ]$. Thus, by setting the reflection phase as $\varpi _{k,n}^{{\rm{d,opt}}}$ for $n\in \{1,\cdots,{{N_{{\rm{S,h}},k}}}\}$, we have
\begin{equation}\label{eq-f-d-limit}
	\begin{aligned}
		\mathop {\lim }\limits_{{{N_{{\rm{S,h}},k}}} \to \infty } \frac{{f_{\max}\left( {{{\bf{\Gamma }}_k}} \right)}}{{{{N_{{\rm{S}},k}}}}} &= \mathop {\lim }\limits_{{{N_{{\rm{S,h}},k}}} \to \infty } \frac{1}{{{{N_{{\rm{S,h}},k}}}}} { \sum\limits_{n = 1}^{{N_{{\rm{S,h}},k}}} {{e^{j\left( {\varpi _{k,n}^{{\rm{d}},{\rm{opt}}} - \varpi _{k,n}^{{\rm{c}},{\rm{opt}}}} \right)}}}}\\
		&= \frac{{{2^b}}}{{2\pi }}\int_{ - \frac{\pi }{{{2^b}}}}^{\frac{\pi }{{{2^b}}}} {e^{jX}dX} = {\rm sinc} (\frac{\pi }{{{2^b}}}),
	\end{aligned}
\end{equation}
which completes the proof.

\section{}\label{App:B}
\textcolor{black}{Applying the exhaustive search scheme for \eqref{eq-RIS_div_orig}, the search complexity is given by
\begin{equation}\label{eq-Comp_ex_1-App}
	\left( {\begin{array}{*{20}{c}}
			K\\
			{s - 1}
	\end{array}} \right) = \frac{{K!}}{{\left( {s - 1} \right)!\left( {K - s + 1} \right)!}}.
\end{equation}
In each search, the computational complexity for the Euclidean norm operation and ${\bf A}^H{\bf A}$ are ${\mathcal O}( {{s^3}} )$ and ${\mathcal O}( {{s^2}{N_{\rm{R}}}} )$, respectively. Considering that $s$ is much less than $N_{\rm R}$ in hybrid mmWave systems, ${\mathcal O}( {{s^3}} )$ can be neglected and the computational complexity in each search is dominated by ${\mathcal O}( {{s^2}{N_{\rm{R}}}} )$. Therefore, the total computational complexity for the exhaustive search is
\begin{equation}\label{eq-Comp_ex_2-App}
	C_1={\mathcal O}\left( {\frac{{{s^2}{N_{\rm{R}}}K!}}{{\left( {s - 1} \right)!\left( {K - s + 1} \right)!}}} \right).
\end{equation}
As a contrast, the maximal search complexity of Algorithm 1 is ${\sum\nolimits_{i = 1}^{s - 1} { {( {K - i + 1} )}} }$ when $\bar k$ is inexistent. In the $i$-th search, the computational complexity is dominated by Line 4 in the Algorithm 1, which is ${\mathcal O}( {{{{{( {i + 1} )}^2}}}{N_{\rm{R}}}} )$. Consequently, the total computational complexity for the greedy search-based algorithm is
\begin{equation}\label{eq-Comp_greedy_1-App}
	\begin{aligned}
		C_2&={\mathcal O}\left( {\sum\nolimits_{i = 1}^{s - 1} \left({ {\left( {K - i + 1} \right)}} {{{\left( {i + 1} \right)}^2}{N_{\rm{R}}}} \right)} \right)\\
		&= {\mathcal O}\left( {{N_{\rm{R}}}\left(\left( {K + 1} \right) {\sum\nolimits_{i = 1}^{s - 1} {{{\left( {i + 1} \right)}^2}}  - \sum\nolimits_{i = 1}^{s - 1} {i{{\left( {i + 1} \right)}^2}} } \right)}\right)
	\end{aligned}.
\end{equation}
Utilizing ${\sum\nolimits_{k = 1}^{n} { {{k}^2}} }=\frac{1}{6}n(n+1)(2n+1)$ and ${\sum\nolimits_{k = 1}^{n} { {k( {k + 1} )^2}} }=\frac{1}{12}n(n+1)(n+2)(3n+5)$ \cite{1980}, \eqref{eq-Comp_greedy_1-App} can be expressed as
\begin{equation}\label{eq-Comp_greedy-2-App}
	\begin{aligned}
		C_2={\mathcal O}&\left( {N_{\rm{R}}}\left(\left( {K + 1} \right)\frac{{s\left( {s + 1} \right)\left( {2s + 1} \right)}}{6} - 1\right.\right.\\
		&\left.\left.- \frac{{\left( {s - 1} \right)s\left( {s + 1} \right)\left( {3s + 2} \right)}}{{12}}\right)\right).
\end{aligned}\end{equation}
Keeping the dominated component in \eqref{eq-Comp_greedy-2-App}, $C_2$ can be approximated by
\begin{equation}\label{eq-Comp_greedy-3-App}
	C_2={\mathcal O}\left( {\frac{{{N_{\rm{R}}}\left( {4K - 3s} \right){s^3}}}{{12}}} \right).
\end{equation}}


\begin{small}

\end{small}

\end{document}